\documentclass[conference]{IEEEtran}
\ifx\pdfoutput\undefined
\usepackage{graphicx}
\else
\usepackage[pdftex]{graphicx}
\usepackage{epstopdf}
\fi
\makeatletter
\def\ps@headings{%
\def\@oddhead{\mbox{}\scriptsize\rightmark \hfil \thepage}%
\def\@evenhead{\scriptsize\thepage \hfil \leftmark\mbox{}}%
\def\@oddfoot{}%
\def\@evenfoot{}}
\makeatother
\usepackage{amssymb}
\usepackage{multirow}
\usepackage[total={7.875in,10.75in},top=0.75in,bottom=1in,left=0.625in,right=0.625in]{geometry}
\usepackage{graphicx}
\usepackage{epsfig}
\usepackage{amsmath}
\usepackage{url}
\usepackage{endnotes}
\usepackage{algorithm}
\usepackage{algorithmic}
\usepackage{multicol}

\newtheorem{theorem}{Theorem}

\begin{document}
\title{On Quantification of Anchor Placement}
\author{\IEEEauthorblockN{Yibei~Ling}
\IEEEauthorblockA{Telcordia Technologies}
\and
\IEEEauthorblockN{Scott~Alexander}
\IEEEauthorblockA{Telcordia Technologies}
\and
\IEEEauthorblockN{Richard Lau}
\IEEEauthorblockA{Telcordia Technologies}}
\maketitle

\begin{abstract}
This paper attempts to answer  
a question:  
for a given traversal area, how 
to quantify the geometric impact of anchor placement on 
localization performance. 
We present a theoretical framework for quantifying 
the anchor placement impact. 
An experimental study, as well as
the field test using a UWB ranging technology, 
is presented.
These experimental results validate the theoretical analysis.
As a byproduct, 
we propose a two-phase localization method (TPLM) and show that 
TPLM outperforms the least-square method in localization accuracy 
by a huge margin. TPLM performs much faster 
than the gradient descent method and 
slightly better than the gradient descent method in 
localization accuracy. 
Our field test suggests that TPLM is more robust against noise 
than the least-square and  gradient descent methods. 
\end{abstract}

\section{Introduction}
Accurate localization is essential for a wide range of 
applications such as mobile ad hoc networking, cognitive
radio and robotics.
The localization problem has many variants that reflect
the diversity of operational environments.
In open areas, GPS
has been considered to be the localization choice. 
Despite its ubiquity and popularity, 
GPS has its Achilles' heel that 
limits its application scope under 
certain circumstances: GPS typically does not work in indoor environments,
and the power consumption of GPS receiver is a major hindrance that precludes
GPS applications in resource-constrained sensor networks.

To circumvent the limitations of GPS,
acoustic/radio-strength and Ultra-wideband (UWB)
ranging technologies have been proposed \cite{Fontana2004,Patwari2005,Gezici2005}.
Measurement technologies include (a) RSS-based (received-signal-strength), (b)
TOA-based (time-of-arrival), and (c) AoA-based (angle-of-arrival). 

RSS-based (received-signal-strength) technologies are most popular 
mainly due to the ubiquity of WiFi and cellular networks.
The basic idea is to translate RSS into distance estimates. 
The performance of RSS-based measurements
are environmentally dependent, due to 
shadowing and multipath effects \cite{Patwari2005,Stoleru2004,Bishop2010}.   
TOA-based technologies such as UWB and GPS measures the propagation-induced time
delay between a transmitter and a receiver. The 
hallmark of TOA-based technologies is 
the receiver's ability to accurately deduce 
the arrival time of the line-of-sight (LOS) signals. 
TOA-based approaches excel RSS-based ones in both the ranging accuracy and 
reliability. In particular, 
UWB-based ranging technologies appear to be promising for indoor positioning as
UWB signal can penetrate most building materials 
\cite{Fontana2004,Patwari2005,Gezici2005}. AoA-based technologies 
measure angles of the target node perceived 
by anchors by means of an antenna array using either RRS or TOA measurement 
\cite{Patwari2005,Gezici2005,Bishop2007}. 
Thus AoA-based approaches are viewed as a variant of RRS or TOA technologies.

Rapid advances in IC fabrication and RF technologies
make possible the deployment of large scale power-efficient sensor networks.  
The network localization problem arises  
from such needs \cite{Tseng2007,Gao2009}.
The goal is to locate all nodes in the network
in which only 
a small number of anchors know their precise positions initially.
A sensor node (with initially unknown position) measures 
its distances to three anchors, and then determines its position. 
Once the position of a node is determined, then the node becomes a new anchor. 
The network localization has two major challenges:
1) cascading error accumulation and 2)
insufficient number of initial anchors and initially skewed anchor distribution. 
The first challenge is addressed by utilizing optimization techniques 
to smooth out error distribution. The second challenge is addressed by using 
multihop ranging estimation techniques \cite{Whitehouse2005,Whitehouse2006}.  

Centralized optimization techniques for solving network localization
include semidefinite programming by So and Ye \cite{So2004} 
and second order cone programming (SOCP) 
relaxation by Tseng \cite{Tseng2007}. 
Local optimization techniques are realistic in practical settings. 
However, they could induce flip ambiguity that deviates significantly
from the ground truth \cite{Gao2009}. To deal with flip ambiguity, 
Eren {\it et. al.} \cite{Eren2004}
ingeniously applied graph rigidity theory to
establish the unique localizability condition. They showed that
a network can be uniquely localizable iff its
grounded graph is globally rigid.
The robust quadrilaterals algorithm by 
Moore et al. \cite{Moore2004} achieves the network localizability 
by gluing locally obtained quadrilaterals, thus 
effectively reducing the likelihood of flip ambiguities. 
Kannan et al. \cite{Kannan2009,Kannan2010} 
formulate flip ambiguity problem to facilitate robust sensor network localization. 

Lederer et al. \cite{Gao2008,lederer08connectivity-j} and 
Priyantha et al. \cite{Priyantha2003} 
revolve around anchor-free localization problem in which none of 
the nodes know their positions. The goal is to construct a 
global network layout by using network connectivity. 
They develop an algorithm for 
constructing a globally rigid Delaunay complex for 
localization of a large sensor network with complex shape.
Bruck, Gao and Jiang \cite{Gao2009} establish 
the condition of network localization using the local angle information. 
They show that embedding a unit disk graph 
is NP-hard even when the angles between adjacent edges are given. 

Patwari et al. \cite{Patwari2005} 
used the Cramer-Rao bound (CRB) to establish performance bounds for
localizing stationary nodes in sensor networks under different path-loss exponents. 
Dulman et al. \cite{Dulman2008} propose an
iteration algorithm for the placement of 
three anchors for a given set of stationary nodes: 
in each iteration, the new position of one chosen anchor 
is computed using the noise-resilience metric. 
Bishop et al. \cite{Bishop2007,Bishop2007a,Bishop2009,Bishop2010}
study the geometric impact of anchor placement 
with respect to one stationary node.
Bulusu et al. proposed adaptive anchor 
placement methods \cite{Bulusu2001}. 
Based on actual localization
error at different places in the region, 
their algorithms can empirically determine good places to deploy additional 
anchors.  

Our work primarily focuses on the geometric impact quantification 
of an anchor placement over a traversal area. 
As a byproduct, it also forms the basis for optimal anchor selection
for mitigating the impact of measurement noise. 
To our best knowledge, only a few papers 
\cite{Dulman2008,Bulusu2001,Bishop2007,Patwari2005,Bishop2007a,Bishop2009,Bishop2010}
mentioned about the geometric effect of anchor placement 
but under a different context from this paper. The idea of quantifying
the geometric impact of anchor placement on localization accuracy over a traversal area
has not been considered before.   

The rest of this paper is structured
as follows.  Section~2 shows examples
to illustrate the anchor placement impact 
on localization accuracy. 
Section~3 proposes a method for quantifying 
the anchor placement impact. 
Section~4 presents a two-phase localization method
Section~5
conducts a simulation study to
validate the theoretical results. Section~6 presents the field study using
the UWB ranging technology. Section~7 concludes this paper.
%\vspace{-1mm}
\section{Effect of Anchor Placement}\label{sect:effectofap}
Throughout this paper we use the term {\it anchor}
to denote a
node with known position. The positions of anchors 
are stationary and available to each mobile node (MN hereafter) 
in a traversal area \cite{Patwari2005}.
The goal is to establish the position of a MN through ranging measurements 
to available anchors. 
The problem is formulated as follows: Let
$p_i=\!(x_{i},y_{i}),\ (1 \leq i \leq m)$ denote the 
known position of the $i^{th}$ anchor, and $p$ the actual position
of the MN of interest. The distance between $p$ and $p_i$ is thus expressed as 
$d_i=d(p,p_i)\!=\!\sqrt{(x-x_i)^2+(y-y_i)^2}$.
In practical terms, the obtained distance measurement is affected by measurement 
noises.  
Thus the obtained distance is as 
$\widehat{d_{i}}=d_i+\epsilon_i$ where $d_i$ the true
distance, and $\epsilon_i$ is widely assumed to be a 
Gaussian noise with zero mean and variance $\sigma^2_i$
\cite{Zhou2004,Youssef2007,Bishop2007a,Langendoen2005,Dulman2008,Patwari2005,Gezici2005}. 
Define a function 
$$f(x,y)=\sum_{i=1}^m \left((x-x_i)^2+(y-y_i)^2-\widehat{d_i}^2\right)^2,$$
where $m$ is the number of accessible anchors, 
$(x_i,y_i)$ is the position of the $i^{th}$ anchor, 
and $\widehat{d_i}$ noisy ranging between 
the $i^{th}$ anchor and the MN. The localization problem 
\cite{Patwari2005,Dulman2008,Langendoen2005} is formulated as
\begin{align} \label{equ:equ1}
\min\limits_{(x,y)\in \Re^2} f(x,y), \Re \mbox{ is the real number set}.
\end{align} 
(\ref{equ:equ1}) presents a nonlinear optimization problem that can be 
solved by a number of algorithms. This paper uses the gradient 
descent method (GDM hereafter) 
for its simplicity \cite{Arfken2005}. 
GDM is based on an intuitive idea that 
if $f(x^{(0)},y^{(0)})$ is differentiable in the 
vicinty of $(x^{(0)},y^{(0)})$, 
then $f((x^{(0)},y^{(0)})$ decreases fastest in the 
direction of the negative gradient 
$-\nabla f(x^{(0)},y^{(0)})$. 
\begin{align} \label{equ:gdm}
\begin{pmatrix}
x^{(i+1)} \\
y^{(i+1)} 
\end{pmatrix}
=\begin{pmatrix}
x^{(i)} \\
y^{(i)} 
\end{pmatrix} 
-\eta 
\begin{pmatrix}
\frac{\partial f(x,y)}{\partial x} \\
\frac{\partial f(x,y)}{\partial y} 
\end{pmatrix},
\end{align}
where $\eta$ refers to the iteration step size, 
$i$ denotes the $i$th iterations. 
One starts with an initial value $(x^{(0)},y^{(0)})$, then applies (\ref{equ:gdm}) 
iteratively to reach a local minimum $(\hat{x},\hat{y})$, {\it i.e.}, 
$f(\hat{x},\hat{y})\!\leq\!f(x^{(i)},y^{(i)}),\ 1\leq i \leq n$. 

There are three issues associated with GDM: 
1) initial value; 2) iteration step size and 3) convergence rate.
The convergence rate could be sensitive to the initial value as well as the iteration 
step size. While in general there is no theoretical 
guidance for selecting the initial value  
in practice, 
the value obtained by a linearized method is often chosen heuristically as the initial value. 
This nonlinear optimization problem in (\ref{equ:equ1}) 
can be linearized by the least-square method \cite{Langendoen2005,Li2005} as follows:
\begin{align} \label{equ:least-square-method}
(A^{T}A)\begin{pmatrix}
x^{(0)} \\
y^{(0)}
\end{pmatrix}= A^T M, \mbox{ where}
\end{align}
\vspace{-2mm}
\begin{flalign} \label{equ:condition}
A=2\begin{pmatrix}
(x_2-x_1)&(y_2-y_1)\\
(x_3-x_1)&(y_3-y_1)\\
\cdots &\cdots \\
(x_{m}-x_1)&(y_{m}-y_1)
\end{pmatrix},
M=\begin{pmatrix}
\widehat{d}_1^2-\widehat{d}_2^2+r_2^2- r_1^2 \\
\widehat{d}_1^2-\widehat{d}_3^2+r_3^2- r_1^2 \\
\cdots \\
\widehat{d}_1^2-\widehat{d}_m^2+r_m^2- r_1^2
\end{pmatrix},
\end{flalign}
$A^T$ is the transpose of $A$, and
$r_i=\sqrt{x_i^2+y_i^2}, 1\leq i \leq m$.
The least-square method (LSM hereafter)
uses noisy measurements 
to estimate the position of MN $(x^{(0)},y^{(0)})^T$. 

\begin{figure}[htb]
\centering
%\centerline{\psfig{file=hilbert.eps,height=1.8in,width=2.5in}}
\resizebox{2.5in}{1.8in}{\includegraphics{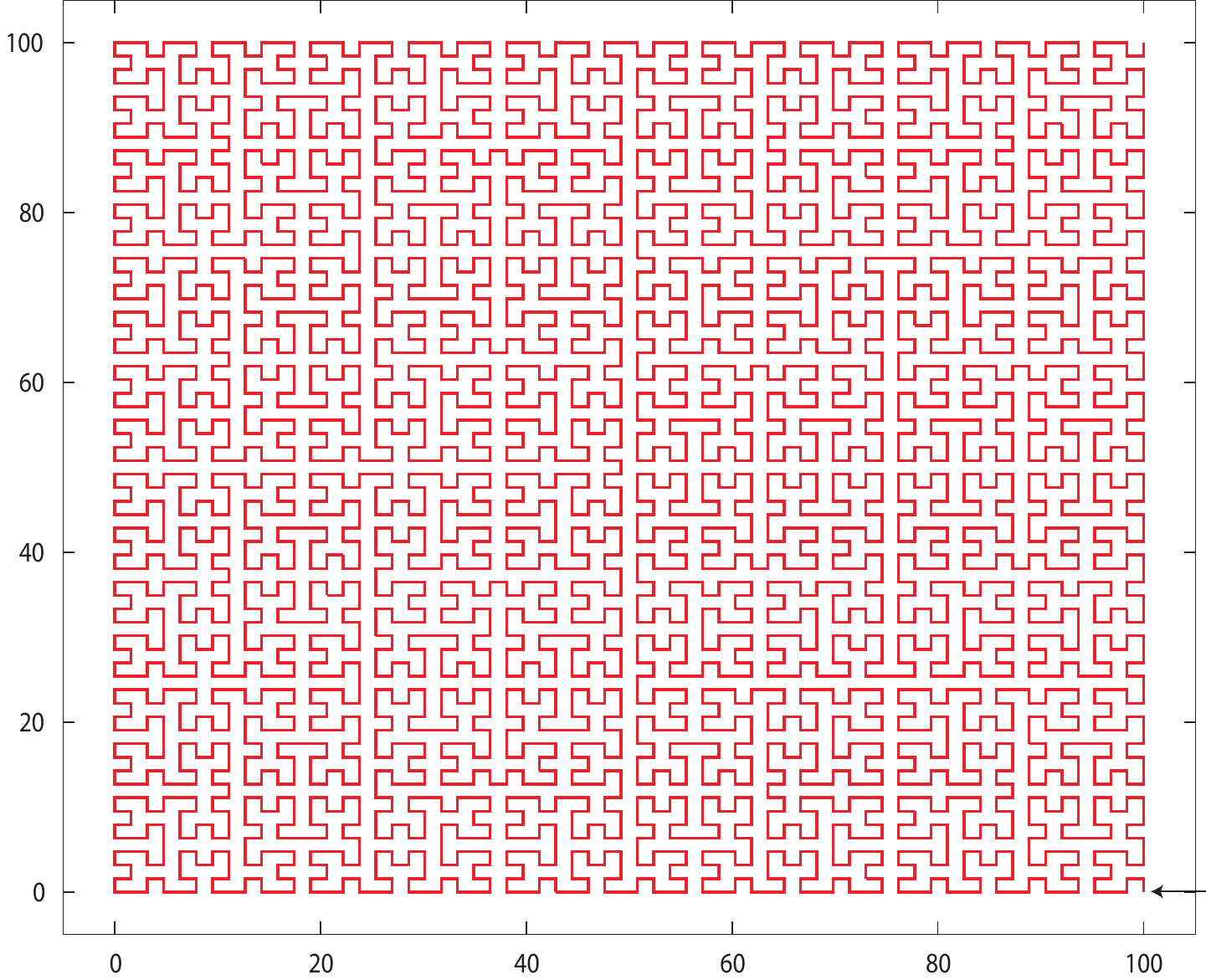}}
\vspace{-2mm}
\caption{Hilbert Trajectory}
\label{fig:hilbert}
\vspace{-2mm}
\end{figure}

To shed light on the anchor placement effect on localization accuracy, 
consider a Hilbert traversal trajectory with 
different anchor placement (AP) setups in 
Table~\ref{tab:anchorm}. 
The Hilbert trajectory is formed by the piecewise connection of $8190$ 
points in an $100\!\times\!100$ region (see 
Fig(\ref{fig:hilbert})). A Hilbert trajectory (HT hereafer) 
has a space-filling property. As a result, 
the localization performance obtained on a HT is a good approximation to
that of the underlying area. 

We conduct an experiment as follows: 
start from the upper-left corner, the MN moves along the HT. 
At each point $p$ (the position of MN), noisy
distance measurements $\widehat{d_i}=d_i+\epsilon_i$ from $p$ to the 
anchors are generated. 
Both LSM and GDM are then used to derive the estimated position. 
To avoid artifacts, 
for a given AP and noise level, the error statistics in 
Table~\ref{tab:anchorm} are obtained by traversing the HT $10$ times. 
Each HT traversal involves the establishment of $8190$ positions using both 
LSM and GDM (GDM uses the estimated position derived from LSM
as its iteration initial value), with a step size of $0.00001$. The termination
condition of GDM is set as $||p^{(i+1)}\!-\!p^{(i)}||_2\!<\!0.001$ 
and the maximum number of iterations is set as $100$. 

\begin{table}[htbp]
\centering
\caption{Anchor placement \& localization accuracy}
\label{tab:anchorm}
\vspace{-2mm}
\begin{tabular}{c|c||c|c|c} \hline
\multicolumn{5}{c}{Least-Square Method} \\ \hline
{positions of anchors} & $\sigma$ &
ave & std & time  \\ \hline
\multirow{2}{*}{$(0,100),(0,0),(100,0)$} & $0.3$ & $0.42$ &$0.26$ & $2.52$  
\\ \cline{2-5}
& $1.0$ & $1.40$ &$0.85$ & $2.54$ \\ \hline
\multirow{2}{*}{$(0,100),(7,50),(3,40)$} & $0.3$ &  
$4.01$ &$3.65$ & $2.51$ \\ \cline{2-5}
&
$1.0$ & $13.45$ &$12.15$ & $2.57$ \\ \hline 
\multicolumn{5}{c}{Gradient Descent Method}\\ \hline
{positions of anchors} &
$\sigma$ & ave & std  & time \\ \hline
\multirow{2}{*}{$(0,100),(0,0),(100,0)$} 
&$0.3$ & $0.37$ &$0.21$  & $63$ 
\\ \cline{2-5}
&
$1.0$ & $1.22$ &$0.69$ & $81$  
\\ \hline
\multirow{2}{*}{$(0,100),(7,50),(3,40)$} & 
 $0.3$ & $0.76$ &$0.92$ &$275$ 
\\ \cline{2-5}
 & $1.0$ &
$2.94$ &$7.41$ 
& $293$ 
\\ \hline
\multicolumn{5}{c}{Gaussian noise ${\cal N}(0,\sigma^2)$} \\ \hline 
\end{tabular}
\end{table}

The ave and std fields in Table~\ref{tab:anchorm} denote 
average localization error and standard deviation, and the time field 
the elapsed time per HT traversal (seconds). 
Table~\ref{tab:anchorm} shows that AP $(0,100),(0,0),(100,0)$
yields a much better localization accuracy than 
AP $(0,100),(7,50),(3,40)$ for both LSM and GDM.
This indicates, a fortiori, that  
an anchor placement (AP) has a significant bearing on 
localization accuracy over an area.
This observation raises a question: {\it can we quantify the impact 
of an anchor placement (AP) with respect to an area?}
The aim of this paper is to answer this question.

\section{Anchor Placement Effect Quantification}
The section focuses on the effect quantification of anchor placement.  
Our main device is based on the notion of the geometric dilution 
of precision (GDOP) \cite{Lee1975,LeeHB1975,Patwari2005,Dulman2008}. 
We begin with the anchor-pair GDOP function as follows:

\begin{theorem} \label{the:gdop}
Let $p_i$ and $p_j$ be the positions of anchor pair $i$ and $j$, and 
$d_i$ and $d_j$ be the actual distances from MN at $p=(x,y)$ to the anchors $i,j$, 
then the geometric dilution of precision at $p$ with respect to $p_i,p_j$, 
denoted by the anchor-pair GDP function $g_2(p_i,p_j)(p)$, is 
\begin{align} \label{equ:gdop}
g_2(p_i,p_j)(p)=
\sqrt{\frac{2}
{1-\left(\frac{d_i^2+d_j^2-||p_i-p_j||^2}{2 d_i d_j}\right)^2}},
\end{align}
where $||p_i\!-\!p_j||$ is the distance between $p_i=(x_i,y_i)$ and $p_j=(x_j,y_j)$. 
\end{theorem}

{\it Proof}:
Let $h(p_i,p_j)(p)$ be a matrix defined as 
\begin{flalign} \label{equ:matrixgdop}
h(p_i,p_j)(p)&\!=\!\begin{pmatrix}
\sin(\alpha)& \cos(\alpha) \\
\sin(\beta)& \cos(\beta)
\end{pmatrix} \\ \nonumber
&=\!\begin{pmatrix}
\frac{(x-x_i)}{\sqrt{(x-x_i)^2+(y-y_i)^2}}, &
\frac{(y-y_i)}{\sqrt{(x-x_i)^2+(y-y_i)^2}} \\
\frac{(x-x_j)}{\sqrt{(x-x_j)^2+(y-y_j)^2}}, &
\frac{(y-y_j)}{\sqrt{(x-x_j)^2+(y-y_j)^2}}
\end{pmatrix},
\end{flalign}
where $(\sin(\alpha),\cos(\alpha))$
and $(\sin(\beta),\cos(\beta))$ denote
the direction cosines from $p$ to anchors 
at $p_i$ and $p_j$. 
The anchor-pair GDOP function $g_2(p_i,p_j)(p)$ is 
\begin{flalign}
g_2(p_i,p_j)(p)=\sqrt{tr((h^Th)^{-1})},
\end{flalign}
where $T/tr$ denotes the transpose/trace of a matrix,
and $A^{-1}$ refers to the inversion of $A$. A simple
manipulation obtains $det(h^Th)=\sin^2(\beta-\alpha)$ and 
\begin{align} \label{equ:ggmm}
g_2(p_i,p_j)(p)= \sqrt{tr((h^Th)^{-1})}
=\sqrt{\frac{2}{\sin^2(\beta-\alpha)}}.
\end{align}  
Substituting 
$\sin^2(\beta-\alpha)\!=1-\!(\dfrac{d_i^2+d_j^2-||p_i-p_j||^2}{2
d_i d_j})^2$ into (\ref{equ:ggmm}) yields (\ref{equ:gdop}). \hfill $\square$

Theorem~\ref{the:gdop} asserts that 
$\beta-\alpha$ degree separation represents the 
effect of anchor pair at $p_i, p_j$ on the localization accuracy at $p$. 
It implies that $\beta-\alpha=\pi/2$ degree 
separation ($g_2(p_i,p_j)(p)=\sqrt{2}$) are best in localization
accuracy, whereas $\beta-\alpha=0$ degree separation (colinear) 
($g_2(p_i,p_j)(p)=\infty$) are worst.   
The smaller the $g_2(p_i,p_j)(p)$, the better 
the accuracy of localization. This is a well-known result also obtained in 
\cite{Patwari2005,Bishop2007a,Bishop2009,Bishop2007,Dulman2008}.

We now extend the anchor-pair GDOP function into the multi-anchor GDOP function.
Let $p_1,\cdots,p_m$ be a set of positions of anchors accessible to MN at $p$. 
The multi-anchor GDOP function $g_m(p_1,\cdots,p_m)(p)$
is linked to the anchor-pair GDOP function as follows: 
\begin{align} \label{equ:tomograph}
g_m(p_1,\cdots,p_m)(p)=\min_{\substack{1 \leq i,j \leq m \\
i \not=j}}g_2(p_i,p_j)(p)
\end{align}

We call an anchor pair $(p_i,p_j)$ the {\it optimally selected anchor pair} (OSAP) if 
$g_2(p_i,p_j)(p)$ has a minimum value among all anchor pairs from 
from $p_1,\cdots,p_m$.
(\ref{equ:tomograph}) shows that $g_m(p_1,\cdots,p_m)(p)$ is determined 
by the OSAP, which in turn varies with $p$ and accessible anchors.  

\begin{figure}[hbt]
\vspace{-4mm}
%\centerline{\psfig{file=lvt3.eps,height=1.5in,width=3.3in}}
\resizebox{3.3in}{1.5in}{\includegraphics{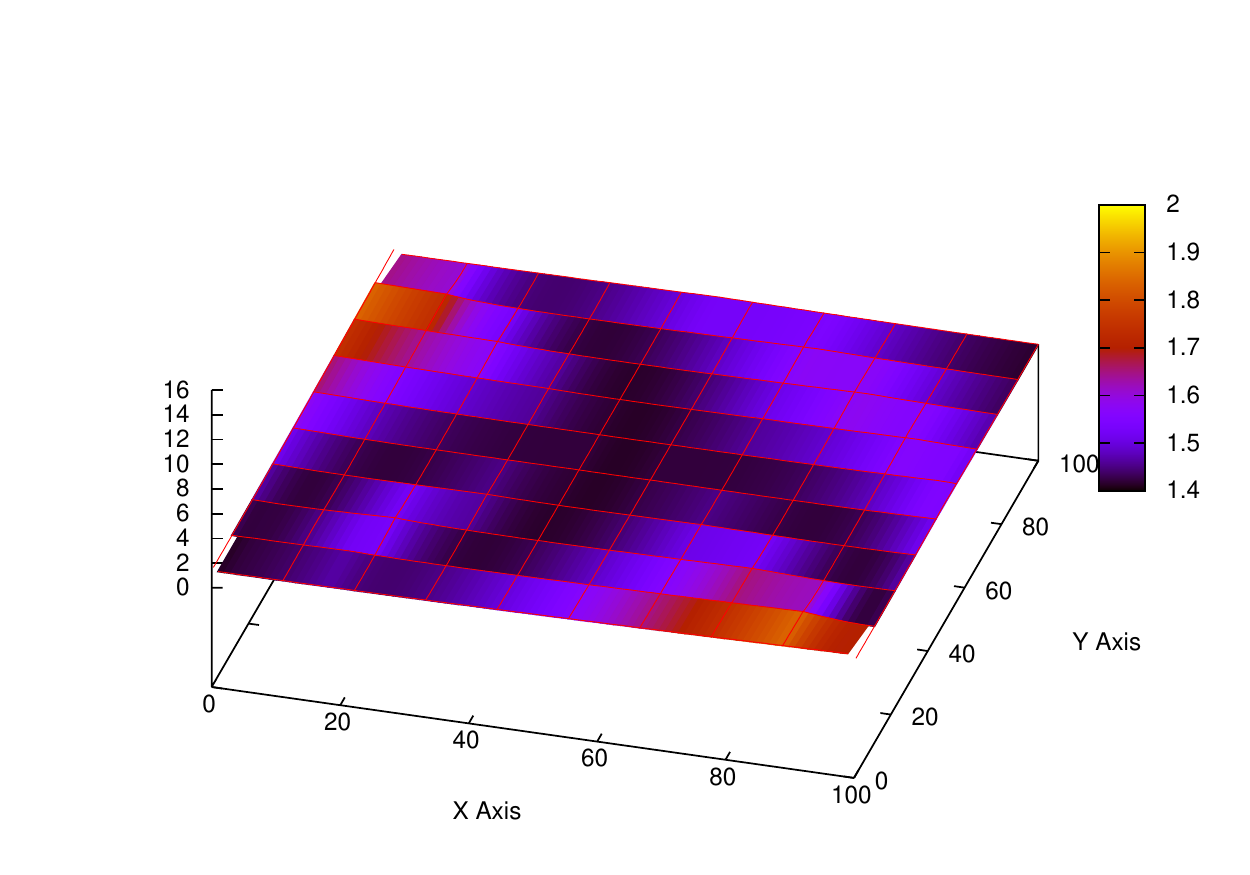}}
\vspace{-2mm}
\caption{{\small LVT of anchor placement $(0,100),(0,0),(100,0)$}}
\label{fig:lvt3}
\vspace{-3mm}
%\centerline{\psfig{file=lvt3a.eps,height=1.5in,width=3.3in}}
\resizebox{3.3in}{1.5in}{\includegraphics{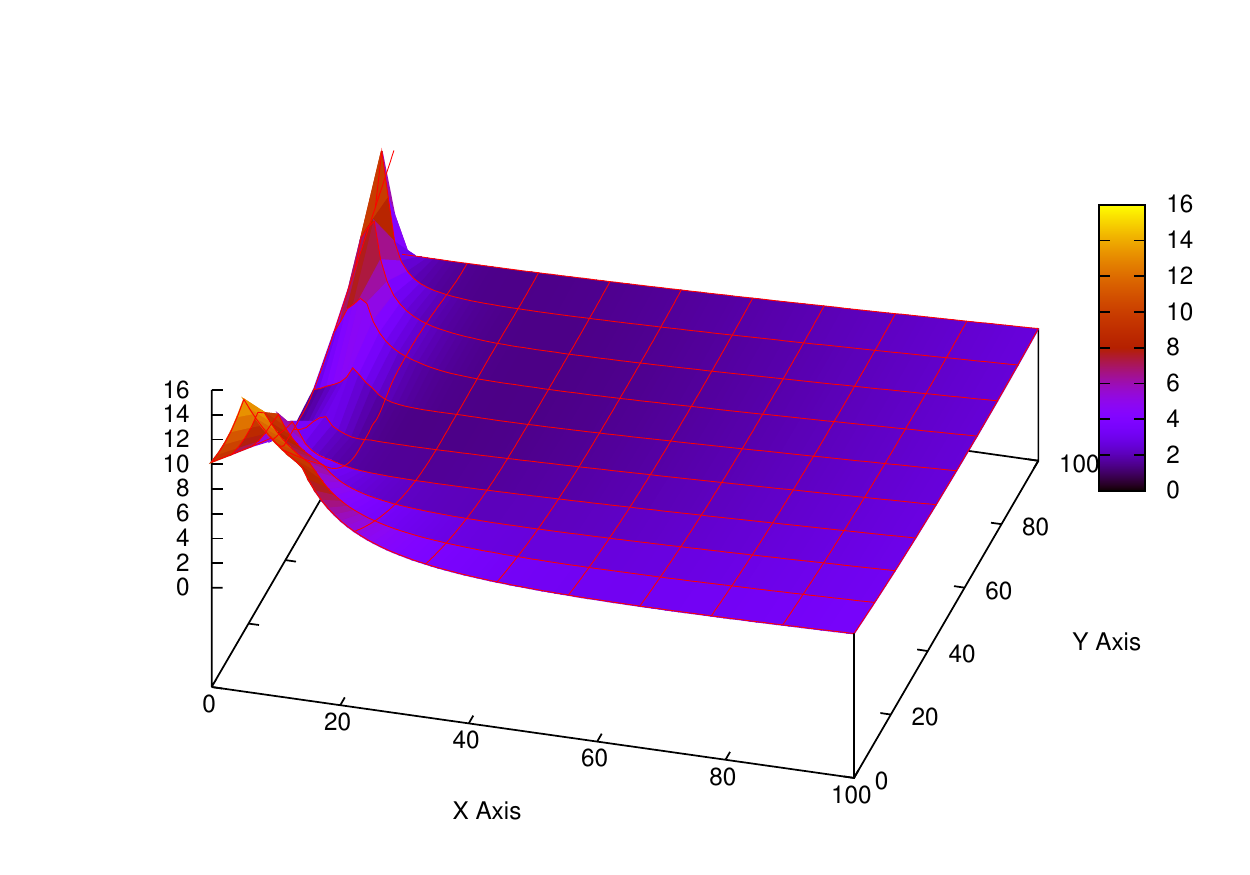}}
\vspace{-2mm}
\caption{{\small LVT of anchor placement $(0,100),(7,50),(3,40)$}}
\label{fig:lvt3a}
\vspace{-2mm}
\end{figure}

Figs(\ref{fig:lvt3})-(\ref{fig:lvt3a}) visualize 
$g_3(\cdots)(p)$ function under the 
two anchor placements over the $100\times 100$ region. 
A three-dimensional graph in 
Figs(\ref{fig:lvt3})-(\ref{fig:lvt3a}),
which is called the {\it least vulnerability tomography} (LVT),
geometrizes the anchor placement effect: 
the LVT elevation has an implication: when in a trough area, the noise 
has less impact on localization accuracy than when 
in a peak area. 
Fig(\ref{fig:lvt3}) 
shows that the LVT of AP $(100,0),(0,0),(0,100)$ has terrain waves 
with a elevation variation from $\sqrt{2}$ to $2$. In contrast, 
the LVT of AP $(0,100),(7,50),(3,40)$ shown in Fig(\ref{fig:lvt3a}) 
is relatively flat for the most part of the region 
but has a dramatic elevation variation from $2$ to $16$ in
the vicinity of $(0,0)$ and $(0,100)$.
Overall, the terrain waves in Fig(\ref{fig:lvt3a}) has a much higher
elevation than that in Fig(\ref{fig:lvt3}). 
This offers an explanation: why AP $(0,100),(0,0),(100,0)$ 
outperforms AP $(0,100),(7,50),(3,40)$ as shown in Table~\ref{tab:anchorm}.

The LVTs in Figs(\ref{fig:lvt3})-(\ref{fig:lvt3a}) induce us 
to extend the GDOP function 
from a point into an area. 
Let $g_m(p_1,\cdots,p_m)(\Omega)$ denote the
average elevation of $g_m(p_1, \cdots, p_m)(p)$ over $\Omega$.
The relation between $g_m(p_1,
\cdots,p_m)(\Omega)$ and $g_m(p_1,\cdots,p_m)(p)$ ($g_m(p_1,\cdots,p_m)(x,y)$) becomes
\begin{align} \label{equ:area}
g_m(p_1,\cdots,p_m)&(\Omega) =
 \dfrac{1}{|\Omega|}\int_{\Omega}
g_m(p_1,\cdots,p_m)(x,y) dx\,dy 
\end{align}

Similarly, 
the notion of $g_m(p_1,\cdots,p_m)(p)$ can be 
extended from a point $p$ to a trajectory $\Gamma$ as 
\begin{flalign}\label{equ:fxaverageline}
 g_m(p_1,\cdots,p_m)(\Gamma) &= \frac{1}{|\Gamma|}
\int_{\Gamma}  g_m(p_1, \cdots,p_m)(x,y) dl
\end{flalign}
The discrete form of (\ref{equ:fxaverageline})
becomes
\begin{align} \label{equ:fxdiscrete}
g_m(p_1,\cdots,p_m)(\Gamma) = \frac{\sum\limits_{1 \leq k \leq n}
g(p_1,\cdots,p_m)(x(t_k),y(t_k))}{n}.
\end{align}

(\ref{equ:area}) provides
a means for quantifying the impact of an AP over an area.
In practice, due to the arbitrariness of the
area boundary and of an anchor placement,
it would be impossible to derive a closed-form expression. 
In this paper, we use {\it Trapezium rule} method 
to compute (\ref{equ:area}). It is done by first 
splitting the area into $10,000$ non-overlapping sub-areas,
then applying the {\it Trapezium rule} on each of these sub-areas.  

\begin{table}[htbp]
\centering
\caption{Anchor Placement Impact Quantification}
\vspace{-2mm}
\label{tab:anchorconfiguration}
\begin{tabular}{|c|c|c|c|} \hline
\multicolumn{4}{|c|}{Anchor Placement} \\  \hline
\multicolumn{4}{|c|}{$p_1=(0,100),p_2=(0,0),p_3=(100,0),p_4=(7,50),p_5=(3,40)$} 
\\ \hline
\multicolumn{4}{|c|}{Anchor Placement Impact over $\Omega$} \\ \hline
 \multicolumn{2}{|c|}{$g_3(p_1,p_2,p_3)(\Omega)$} &
 \multicolumn{2}{|c|}{$g_3(p_1,p_4,p_5)(\Omega)$} \\ \hline
\multicolumn{2}{|c|}{$1.501$} & 
\multicolumn{2}{|c|}{$2.615$} \\ \hline
\multicolumn{4}{|c|} {Anchor Placement Impact over $\Gamma$}  \\ \hline
\multicolumn{2}{|c|}{$g_3((p_1,p_2,p_3)(\Gamma)$} & 
\multicolumn{2}{|c|}{$g_3(p_1,p_4,p_5)(\Gamma)$} \\ \hline
 \multicolumn{2}{|c|}{$1.499$}  &
\multicolumn{2}{|c|}{$2.622$} \\ \hline 
\multicolumn{4}{|c|}{$\Omega$: $100\times 100$ area,$\Gamma$: HT} \\ \hline
\end{tabular}
\vspace{-2mm}
\end{table}

Table~\ref{tab:anchorconfiguration} provides a 
numerical calculation of AP impact over $\Omega$ and over $\Gamma$. 
The AP impact calculated in Table~\ref{tab:anchorconfiguration} 
implies that the localization accuracy over the $AP_1$ is better
than that under $AP_2$ as $g_3(p_1,p_2,p_3)(\Omega)<g_3(p_1,p_4,p_5)(\Omega)$,
which agree with the results obtained in noisy environments in Table~\ref{tab:anchorm}.
Comparing $g_3(p_1,p_2,p_3)(\Omega)$ and $g_3(p_1,p_2,p_3)(\Gamma)$ 
in Table~\ref{tab:anchorconfiguration} 
shows that the difference between them is very small.
However, computation of $g_3(p_1,p_2,p_3)(\Omega)$ takes about $23$ seconds,
as opposed to $0.2$ seconds in computing $g_3(p_1,p_2,p_3)(\Gamma)$.

\section{Two-Phase Localization Algorithm}
\begin{figure}[hbt]
%\centerline{\psfig{file=oaps-compare3-black.eps,height=1.4in,width=3.3in}}
\resizebox{3.4in}{1.4in}{\includegraphics{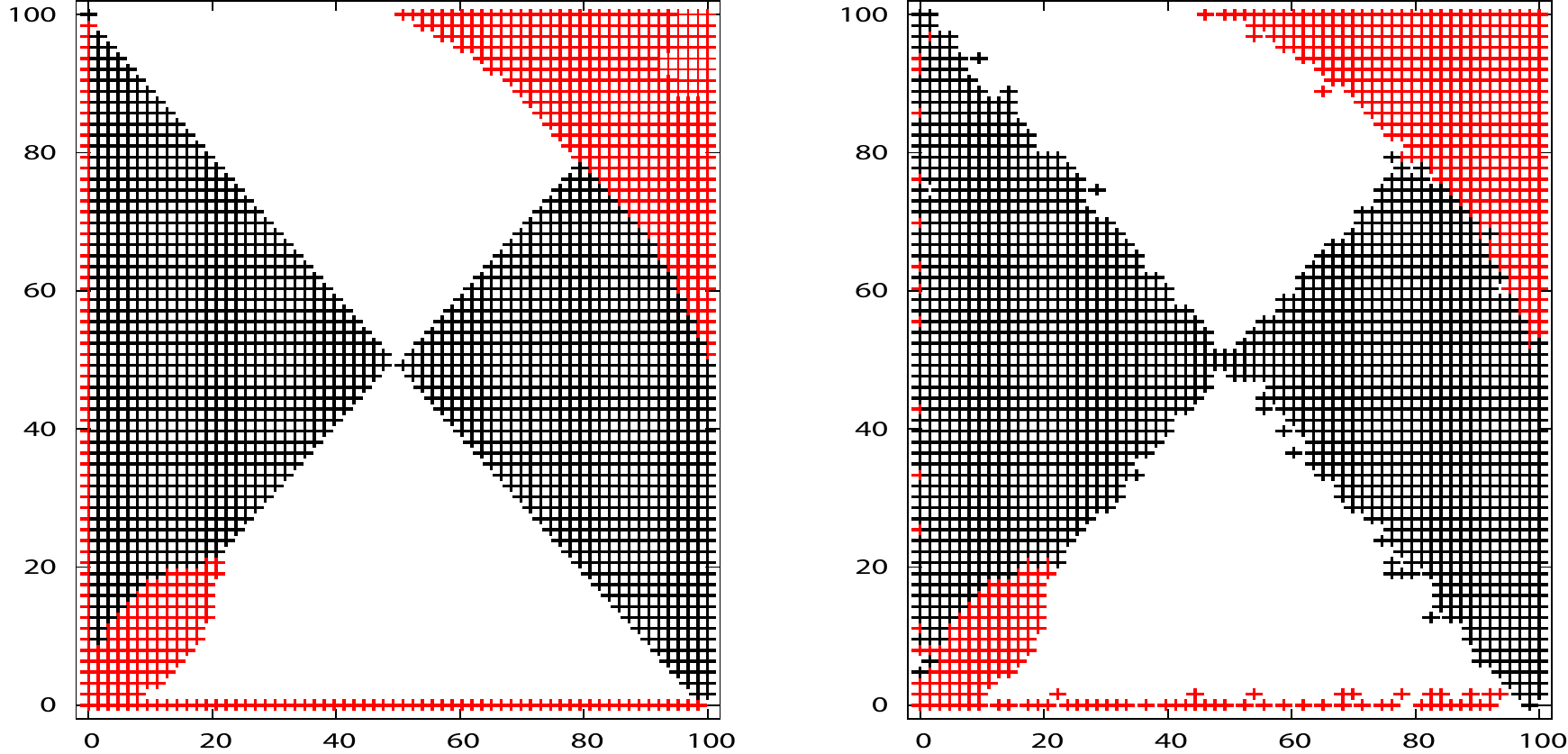}}
\caption{{\small a) noise free b) noise level $\sigma(1.0)$: white area 
by anchor pair at $(0,100)\!,\!(0,0)$, black area by anchor pair at $(0,0),(100,0)$, 
red area by anchor pair at $(0,100)\!,\!(100,0)$}}
\label{fig:oaps} 
\vspace{-2mm} 
\end{figure}

It is clear from Fig(\ref{fig:lvt3}) that 
$g_m(p_1,\cdots,p_m)(p)$ is determined by an OSAP and 
the OSAP varies by area.
Fig(\ref{fig:oaps}) shows the OSAP areas under AP
$(0,100),(0,0),(100,0)$. 
Observe the noise-free scenario in the left-hand side graph in Fig(\ref{fig:oaps})
where each colored area 
represents an anchor pair chosen 
over its two alternatives based on (\ref{equ:tomograph}).
The white colored areas represent the regions in which
the anchor pair at $(0,100),(0,0)$ is chosen. 
The blue colored areas represent the regions where
the anchor pair at $(0,0),(100,0)$ has its geometric 
advantage over its alternative anchor pairs.
The red colored areas represent the regions 
where the anchor pair at $(100,0),(0,100)$ 
is expected to produce the most accurate localization.    
The right-hand side graph in 
Fig(\ref{fig:oaps}) plots 
the OSAP areas under the noise level of $\sigma=1.0$. The presence of noise
makes the borderlines on different colored 
areas rough and unsmooth. This is because that  
borderlines correspond to the isolines by 
two anchor pairs and hence are highly sensitive to noise. 

This observation of Fig(\ref{fig:oaps})
leads to a new two-phase localization algorithm (TPLM) that 
differs significantly from LSM and GDM.
TMPL differs from LSM and GDM. It
proceeds in two phases:
In the first phase called {\it optimal anchor pair selection},
an optimal anchor pair at $(p_i,p_j)$ is identified 
based on (\ref{equ:tomograph}), then the position(s) can 
be estimated by solving 
\begin{flalign}\label{equ:two}
||\widehat{p}-p_i||=\widehat{d_i},\  ||\hat{p}-p_j||=\widehat{d_j}
\end{flalign}
Through a lengthy manuplation, two possible solutions
to (\ref{equ:two}) are obtained as follows
\begin{flalign} \label{equ:matrixtwo} 
&\begin{pmatrix}
\widehat{x} \\ \widehat{y}
\end{pmatrix}=
\begin{pmatrix}
\cos(\theta)& \sin(\theta) \\
\sin(\theta)& \cos(\theta)
\end{pmatrix}
\begin{pmatrix}
u \\
v
\end{pmatrix} + 
\begin{pmatrix}
x_i \\
y_i
\end{pmatrix}   \nonumber  \\
& 
\begin{pmatrix}
\widehat{x} \\
\widehat{y}
\end{pmatrix}=
\begin{pmatrix}
\cos(\theta)& \sin(\theta) \\
\sin(\theta)& \cos(\theta)
\end{pmatrix}
\begin{pmatrix}
u \\
-v
\end{pmatrix} + 
\begin{pmatrix}
x_i \\
y_i
\end{pmatrix}
\end{flalign} 
where $\theta=\arctan(\frac{y_j-y_i}{x_j-x_i})$, $d_{ij}=||p_i-p_j||$, and
\begin{flalign}  \label{equ:tttt}
u=\frac{(d_{ij}^2+\widehat{d_i}^2-\widehat{d_j}^2)}{2 d_{ij}},\ 
v=\sqrt{d_i^2-\left(\frac{d_{ij}^2 + 
\widehat{d_i}^2-\widehat{d_j}^2}{2 d_{ij}}\right)^2}
\end{flalign} 
In the second phase called {\it disambiguation},  the reference point obtained from LSM
is used to single out one from two possible positions (\ref{equ:matrixtwo}).
It is done by choosing one position that has a shorter distance to the reference point. 

\begin{algorithm} 
\caption{Two-phase Localization Method}
\begin{algorithmic}[1]\label{alg:ooo}
\REQUIRE{$p_i=(x_i,y_i), 1\leq i \leq m$: $i^{th}$ anchor's position} 
\REQUIRE{$\bar{d_i}$: distance measurement from $j^{th}$ anchor to MN $p$}
\REQUIRE{$\breve{p}=(\breve{x},\breve{y})$: a reference point obtained via LSM} 
\STATE $g_{min} \leftarrow \infty, p_+\leftarrow(0,0), p_{++}\leftarrow (0,0)$ 
\FOR {$i =1$ to $m-1$}  \label{line:1} 
\FOR {$j=i+1$ to $m$} 
\IF {$g_2(p_i,p_j)(p) <g_{min}$}
\STATE $g_{min}\leftarrow g(p_i,p_j)(p), (p_+,p_{++})\leftarrow (p_i,p_j)$ \label{line:12}
\ENDIF
\ENDFOR
\ENDFOR \label{line:2}
\STATE $d=|p_+-p_{++}|,d_+= |p-p_{+}|,d_{++}=|p-p_{++}|$  \label{line:3}\\
\STATE $\theta=\arctan{\frac{y_{++}-y_{+}}{x_{++}-x_+}},u=\frac{d_+^2-d_{++}^2+d^2}{2d},v=\sqrt{d_+^2-u^2}$ 
\STATE $p_{*}=(\begin{smallmatrix}\cos(\theta) & -\sin(\theta) \\ 
\sin(\theta) & \cos(\theta) 
\end{smallmatrix}) \binom{u}{v} + 
\binom{x^+}{y^+}$
\STATE $p_{**}=(\begin{smallmatrix}\cos(\theta) & -\sin(\theta) \\
\sin(\theta) & \cos(\theta) \end{smallmatrix})\binom{u}{-v}+\binom{x^+}{y^+}$ \label{line:4}
\IF {$|p_*-\breve{p}|< |p_{**}-\breve{p}|$}\label{line:5}
\RETURN $p_*$
\ENDIF
\RETURN $p_{**}$ \label{line:6}
\end{algorithmic}
\end{algorithm}

The code between lines~\ref{line:1}-\ref{line:2} examines
each anchor pair in turn and chooses an optimal anchor pair, 
which  
corresponds to (\ref{equ:tomograph}). The complexity is $\binom{m}{2}$.
Lines~\ref{line:1}-\ref{line:4} 
constitute the basic block of the 
{\it optimal anchor selection phase}.  
The code between lines~\ref{line:5}-\ref{line:6} 
is used to disambiguate the two possible solutions using the reference position 
$\breve{p}$ obtained by LSM. 

\begin{table}[htbp]
\centering
\caption{Two-Phase Localization Method}
\vspace{-1mm}
\label{tab:optimal}
\begin{tabular}{c|c||c|c|c} \hline
{positions of anchors} & $\sigma$ & 
ave & std & time\\ \hline
\multirow{2}{*}{$(0,100),(0,0),(100,0)$} & 
$0.3$ & $0.40$ &$0.22$ &$3.17$\\ \cline{2-5}
& 
$1.0$& $1.32$ &$0.73$ &$3.17$ \\ \hline
\multirow{2}{*}{$(0,100),(7,50),(3,40)$} & 
$0.3$ & $0.70$ &$0.88$ &$3.44$\\ \cline{2-5}
& 
$1.0$ & $2.76$ &$5.67$ & $3.15$\\
\hline 
\multicolumn{5}{c}{Gaussian noise ${\cal N}(0,\sigma^2)$} \\ \hline 
\end{tabular}
\vspace{-1mm}
\end{table}

We present the results of TPLM in Table~\ref{tab:optimal} using the  
same anchor placement setups in Table~\ref{tab:anchorm}.  
In all cases, TPLM gives a significant error 
reduction over LSM with a minor degradation 
as TPLM uses the reference point obtained from LSM.
It is fairly clear that TPLM is an order of magnitude faster than GDM, and 
performs slightly better
than GDM in terms of localization accuracy.  

\section{Simulation Study}
The aim of this section is threefold: first, to conduct
the performance comparison between TPLM, LSM, and GDM in noise environments;
second, to study the impact of noise models; and third, to investigate 
the random anchor placement impact.     
\subsection{Visualizing Noise-induced Distortion}
\begin{figure}[htb]
%\centerline{\psfig{file=lsmrestorehilbert3.eps,height=1.7in,width=3.4in}}
\resizebox{3.4in}{1.7in}{\includegraphics{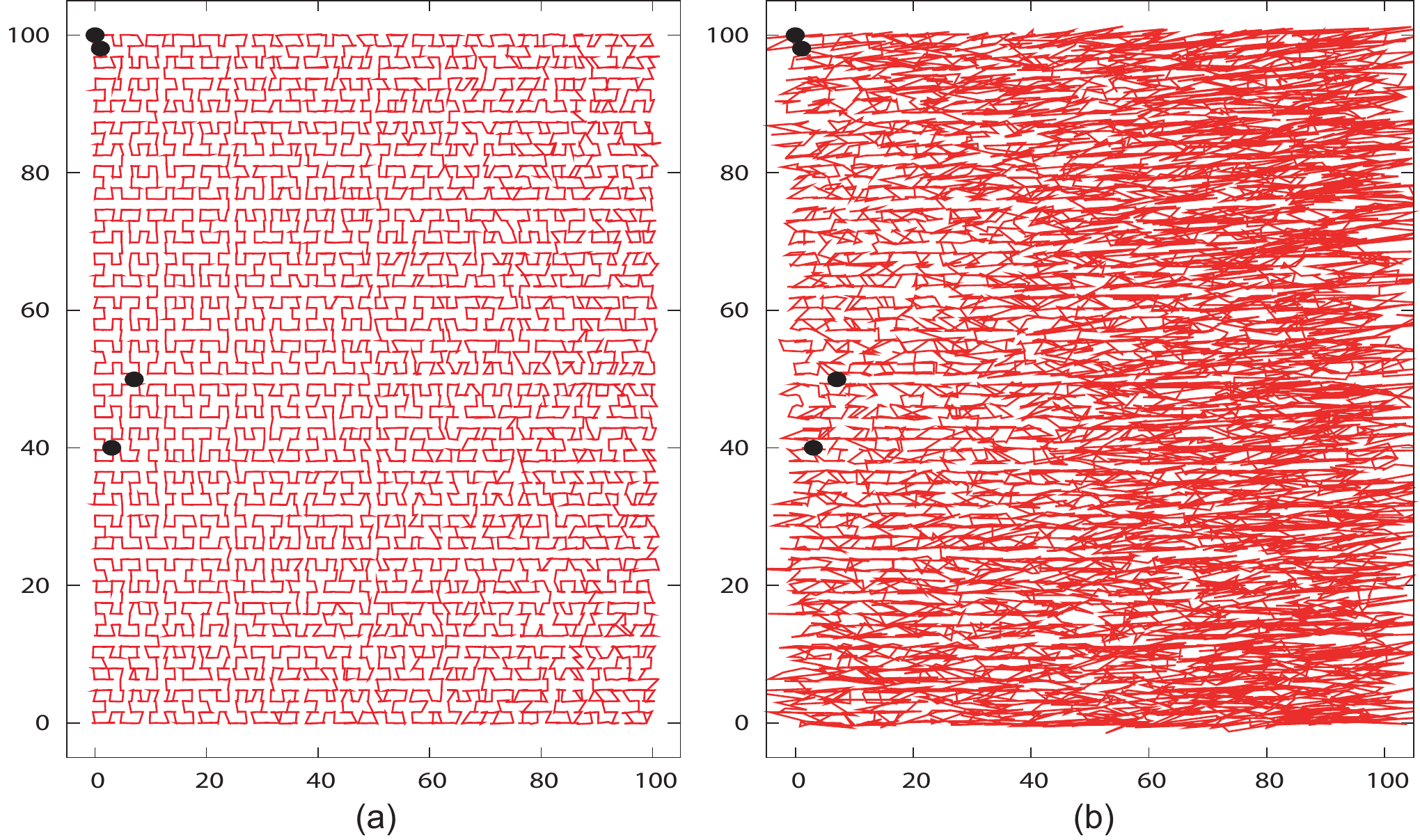}}
\vspace{-3mm}
\caption{{\small restored HT by LSM: a) noise $(0.02)$; b) noise $(0.2)$}}
\label{fig:lsmrestorehilbert1}
\vspace{-1mm}
\end{figure}

\begin{figure}[htb]
\vspace{-1mm}
%\centerline{\psfig{file=gdmrestorehilbert2.eps,height=1.7in,width=3.4in}}
\resizebox{3.4in}{1.7in}{\includegraphics{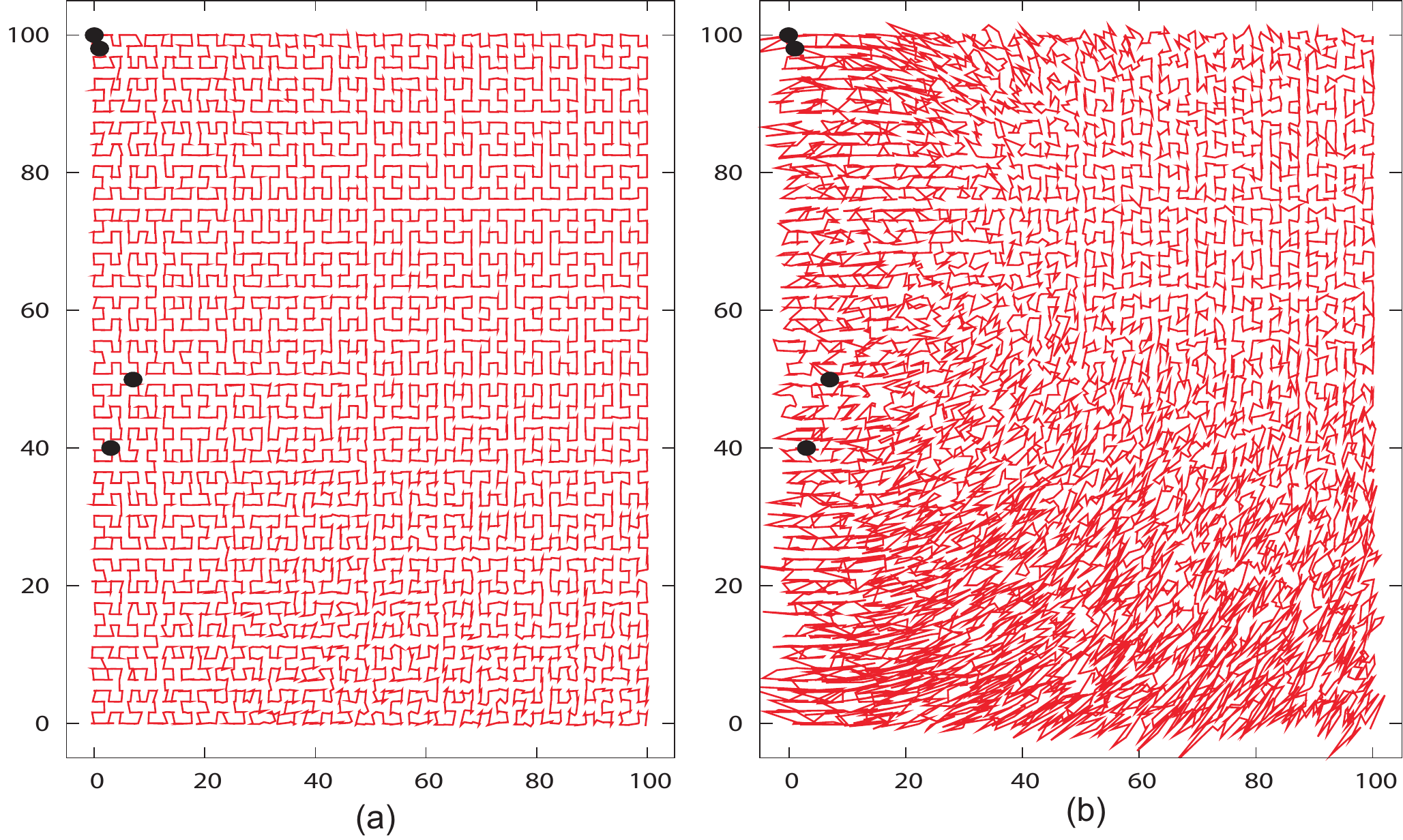}}
\vspace{-3mm}
\caption{{\small restored HT by GDM: a) noise $(0.02)$; b) noise $(0.2)$}}
\label{fig:gdm}
\vspace{-1mm}
\end{figure}

\begin{figure}[htb]
\vspace{-1mm}
%\centerline{\psfig{file=tplmrestorehilbert1.eps,height=1.7in,width=3.4in}}
\resizebox{3.4in}{1.7in}{\includegraphics{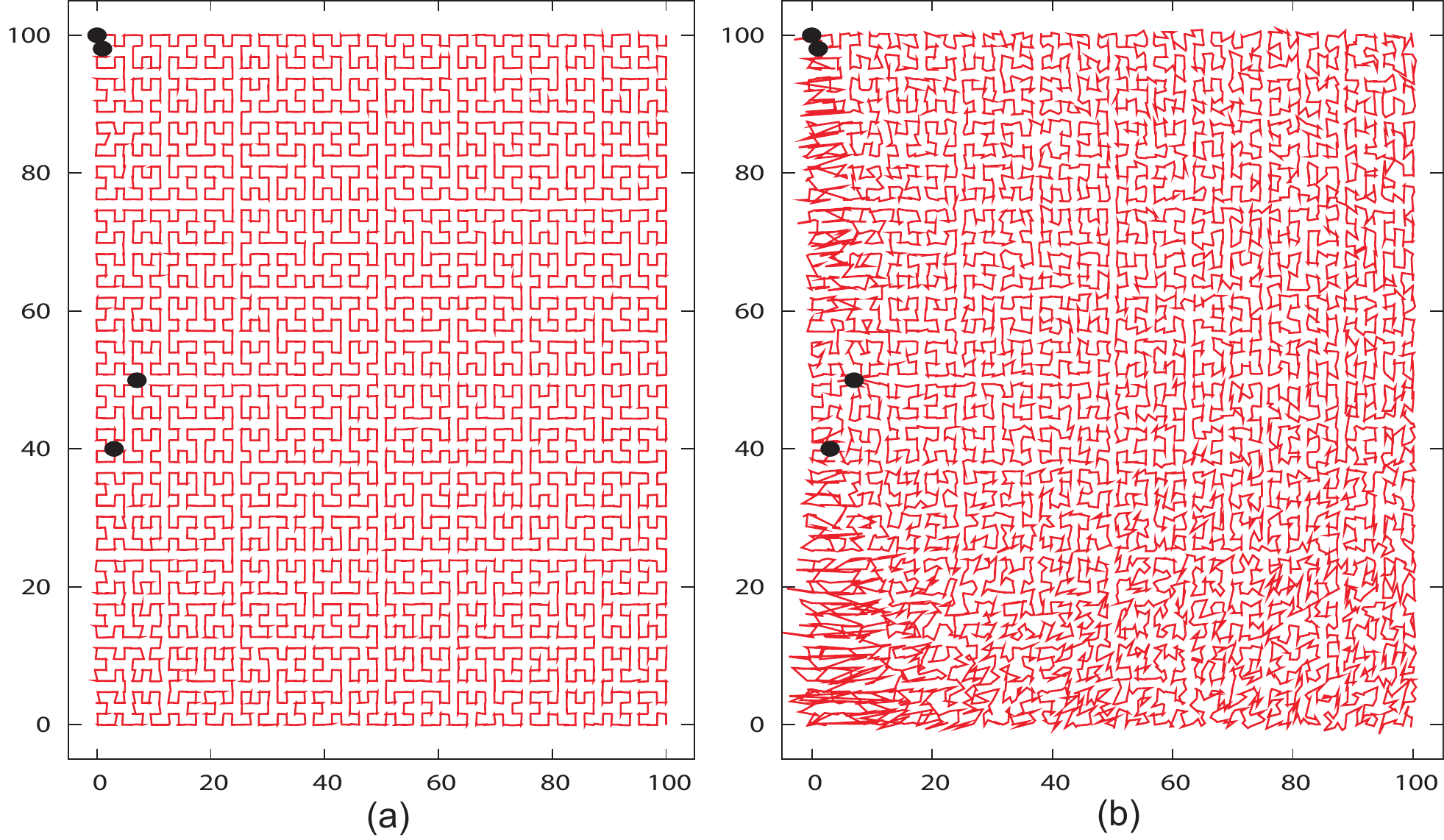}}
\vspace{-3mm}
\caption{{\small restored HT by TPLM: a) noise $(0.02)$; b) noise $(0.2)$}}
\label{fig:tplm}
\vspace{-1mm}
\end{figure}

To visualize the difference among LSM, GDM, and TPLM, we 
plot the restored HT by LSM, GDM, and TPLM under
Gaussian noise model in
Figs(\ref{fig:lsmrestorehilbert1})-(\ref{fig:tplm}) using 
$AP_4$ in Table~\ref{tab:anchorsetup}. Notice that 
anchor positions are plotted as solid black circles.

\begin{table}[htbp]
\centering
\caption{Anchor Placement Setup}
\vspace{-2mm}
\label{tab:anchorsetup}
\begin{tabular}{|c|c|c|c|} \hline
\multicolumn{4}{|c|}{Anchor Position} \\  \hline
\multicolumn{4}{|c|}{$p_1=(0,100),p_2=(7,50),p_3=(0,0)$}\\ 
\multicolumn{4}{|c|}
{$p_4=(3,40),p_5=(100,0),p_6=(1,98)$}\\ \hline
\multicolumn{4}{|c|}{Anchor Placement Setup} \\  \hline
$AP_1$& 
\multicolumn{3}{|c|}{$(0,100),(0,0),(100,0)$} \\ \hline
$AP_2$ & 
\multicolumn{3}{|c|}{
$(0,100),(7,50),(3,40)$} \\ \hline
$AP_3$&  
\multicolumn{3}{|c|}{
$(0,100),(0,0),(100,0),(1,98)$} \\ \hline
$AP_4$ & 
\multicolumn{3}{|c|}{$(0,100),(7,50),(3,40),(1,98)$}   
\\ \hline
\end{tabular}
\end{table}

Visual inspection reveals the apparent perceived difference among the 
restored HTs by 
LSM, GDM and TPLM: when the noise level $\sigma$ is $0.2$, the restored HT  
by LSM becomes completely unrecognizable in Fig(\ref{fig:lsmrestorehilbert1})(b).
While in Fig(\ref{fig:gdm})(b) the upper right portion of the restored HT by 
GDM to some degree preserves 
the hallmark of the HT, but the most part of
the recovered HT  
is severely distorted and barely recognizable. 
The restored curve by TPLM contrasts sharply with those by LSM and GDM in 
its preservation of fine details of the HT for the most part, 
while having minor distortion in the lower- and upper-left corner areas in Fig(\ref{fig:tplm})(b).
Such spatially uneven localization performance can be 
explained by examining Fig(\ref{fig:lvt3a}), which shows that the 
lower- and upper-left areas exactly 
correspond to the LVT peak areas. 
The perceived differences between GDM and TPLM in 
Figs(\ref{fig:lsmrestorehilbert1})-(\ref{fig:tplm}) 
can be quantified as follows: 
GDM has the average error of $0.69$ per HT traversal, while TPLM produces the average  
error of $0.465$.
In addition, GDM takes about $218.48$ seconds per HT traversal, in contrast to  
$2.94$ seconds taken by TPLM.

\subsection{Gaussian Noise vs. Non-Gaussian Noise}
\begin{figure}[htb]
%\centerline{\psfig{file=gdm-tplm-norm-anchors.eps,height=2.5in,width=3.4in}}
\resizebox{3.4in}{2.5in}{\includegraphics{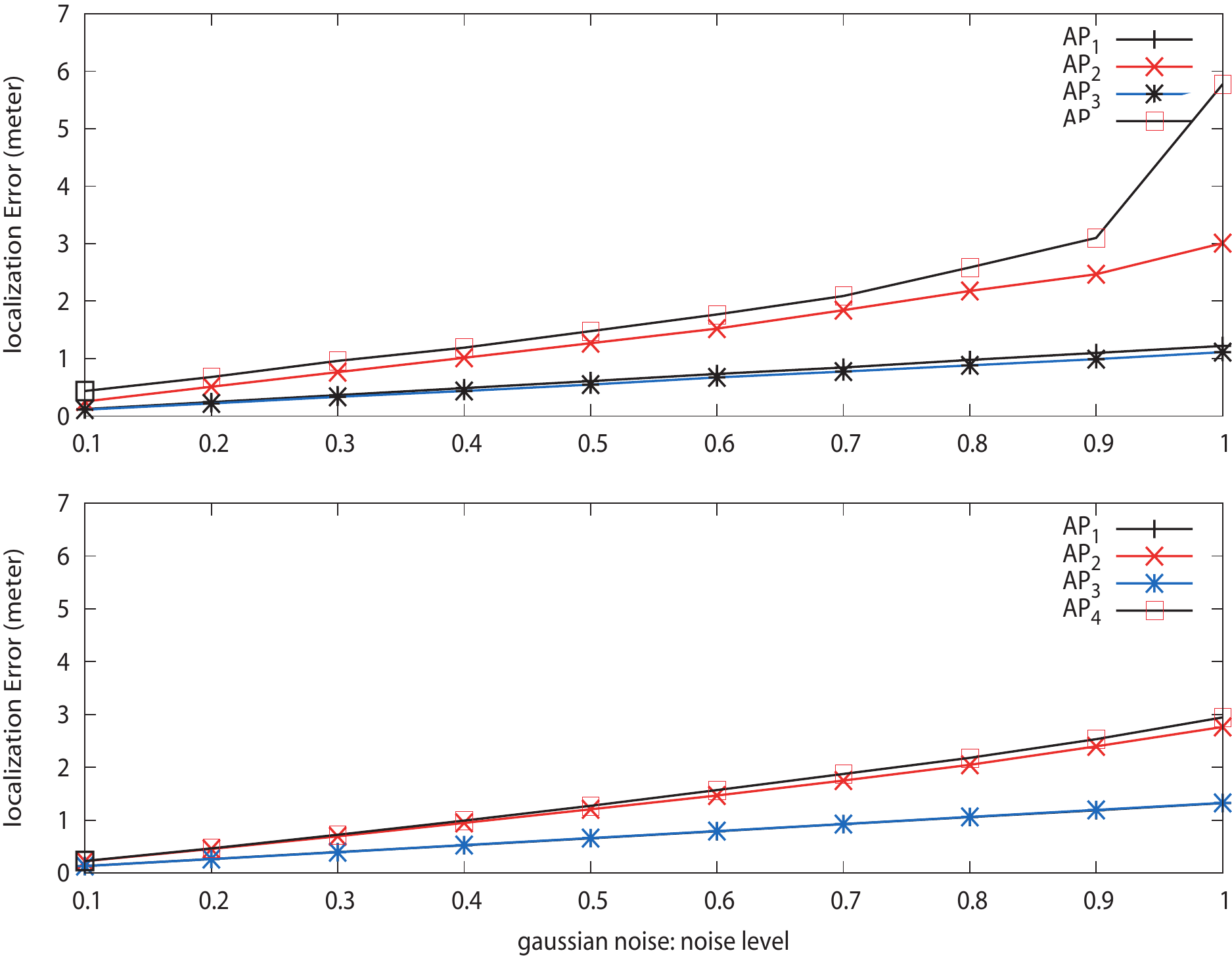}}
\vspace{-2mm}
\caption{{\small Accuracy: (top) GDM; (bottom) TPLM (Gaussian noise)}}
\label{fig:norm-noise}
\vspace{-1mm}
\end{figure}

\begin{figure}[htb]
%\centerline{\psfig{file=gdm-tplm-uniform-anchors.eps,height=2.5in,width=3.4in}}
\resizebox{3.4in}{2.5in}{\includegraphics{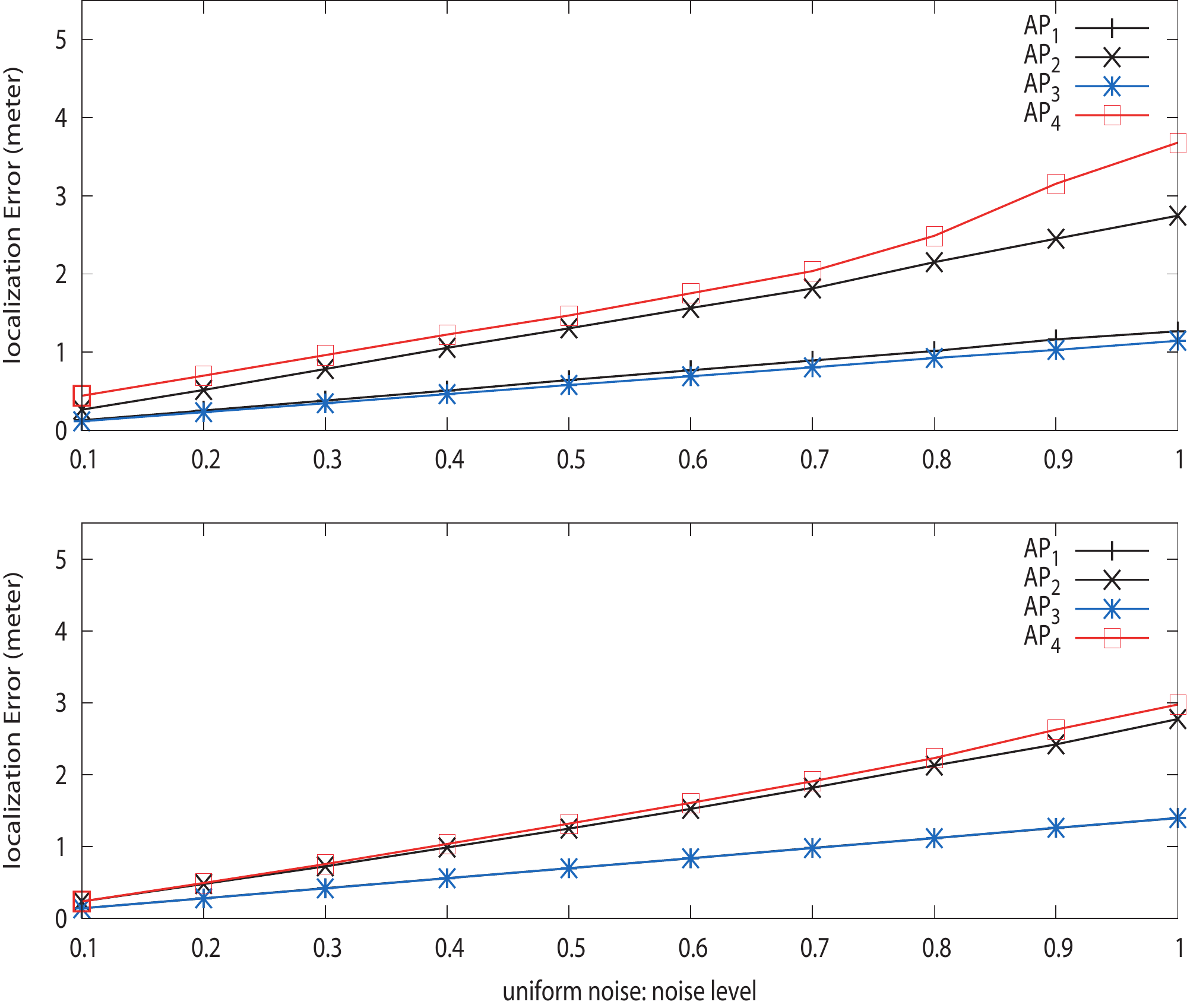}}
\vspace{-2mm}
\caption{{\small Accuracy: (top) GDM; (bottom) TPLM  (uniform noise)}}
\label{fig:uniform-noise}
\vspace{-1mm}
\end{figure}

In this subsection, we will compare the performance of GDM and 
TPLM under Gaussian and uniform noise models,
there are some scenarios where ranging noise may
follow uniform model \cite{Bulusu2001,Liu2008}. 
In the experimental study,
the performance between GDM and TPLM is compared under a same 
noise level with the different noise models.
For Gaussian noise model ${\cal N}(0,\sigma^2)$,
$\sigma$ refers to the noise level in Fig(\ref{fig:norm-noise}), and 
for a uniform noise model $U(-a,a)$, the noise level in Fig(\ref{fig:uniform-noise}) 
is expressed 
as $a^2/3$. The data presented reflects 
the average error of GDM and TPLM over 
$10$ HT traversals under the anchor placements in Table~\ref{tab:anchorconfiguration}.

The graphs in Figs(\ref{fig:norm-noise})-(\ref{fig:uniform-noise})
represent 
the localization error curves of GDM and of TPLM, respectively.
Under both the noise models, TPLM outperforms GDM by huge margins 
under $AP_2$ and $AP_4$.
Both GDM and TPLM  
perform indistinguishably under $AP_1$ and $AP_3$.
While the impact difference between $AP_1$ and $AP_3$ is barely noticed as their 
performance curves are overlapped,
the performance curves between $AP_1$ and $AP_2$ and between $AP_3$ and $AP_4$ are 
clearly separated, thus 
the impact difference $AP_1$ and $AP_2$ and between $AP_3$ and $AP_4$ are evident.  

\begin{figure}[htb]
%\centerline{\psfig{file=gdm-tplm-norm-time.eps,height=2.0in,width=3.4in}}
\resizebox{3.4in}{2in}{\includegraphics{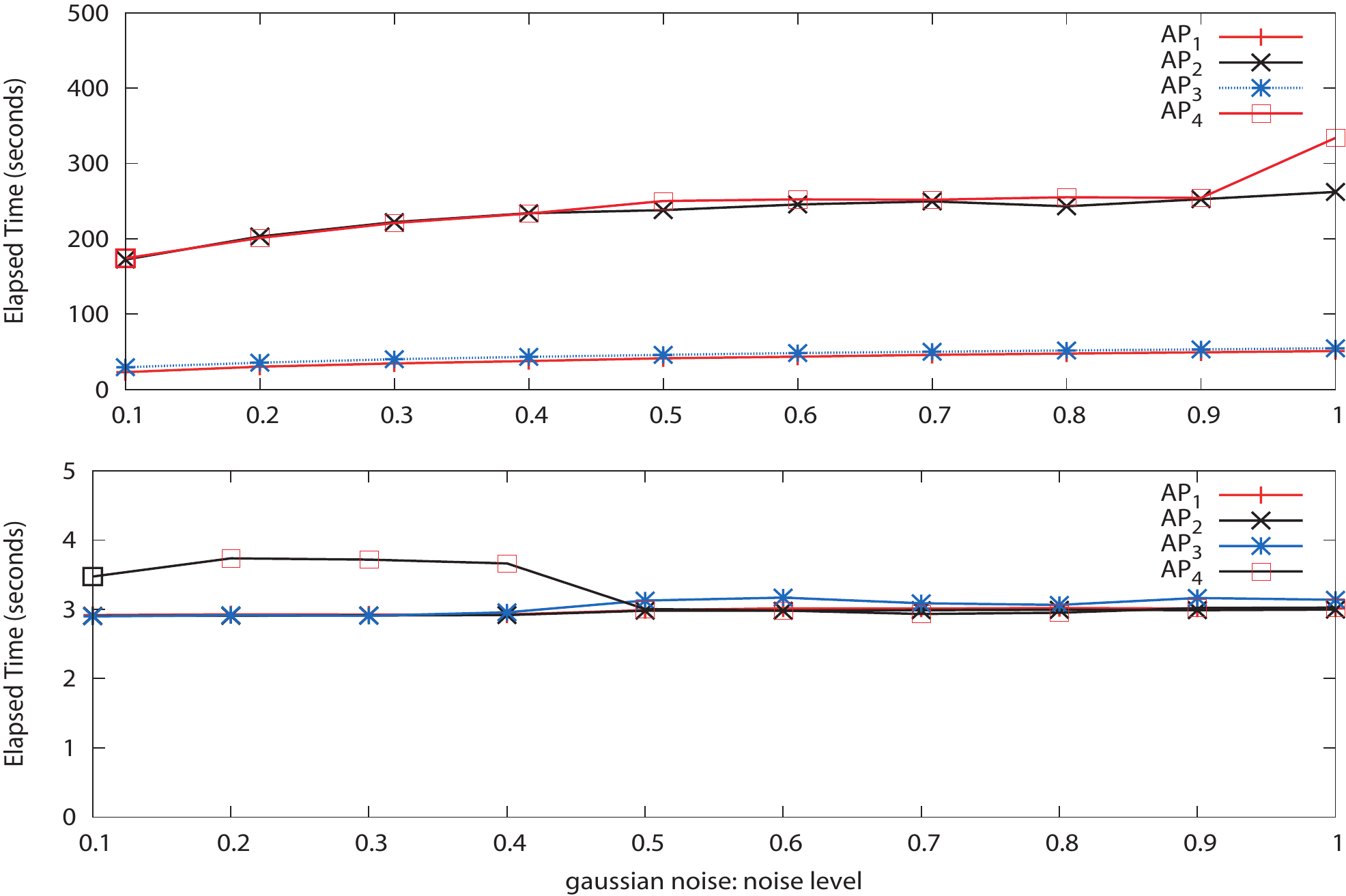}}
\vspace{-2mm}
\caption{{\small Elapsed time: (top) GDM; (bottom) TPLM  (normal noise)}}
\label{fig:norm-noise-time}
\vspace{-1mm}
\end{figure}

The results show that the Gaussian noise has more impact on GDM than the uniform one.  
By comparison, the performance of TPLM is insensitive to noise models.
Fig(\ref{fig:norm-noise-time}) plots   
the elapsed time by GDM and TPLM per HT traversal, showing that 
TPLM is orders of magnitude faster than 
GDM.    

\subsection{Random Anchor Placement}
We study the impact of random anchor placement on the performance of  LSM, GDM and TLPM.
To achieve this, the number of anchors are randomly placed in  
$100\times 100$ and $50 \times 100$ regions. 
For each randomly generated anchor placement (RGAP) in a region, 
the $g_m(p_1,\cdots,p_m)(\Gamma)$ function is calculated, and
the error statistics of LSM, GDM, and TPLM under Gaussian noise level of $0.3$ are 
gathered and compared.

\begin{figure}[htb]
\centering
%\centerline{\psfig{file=gdistribution-uniform-skew.eps,height=2.5in,width=3.4in}}
\resizebox{3.4in}{2in}{\includegraphics{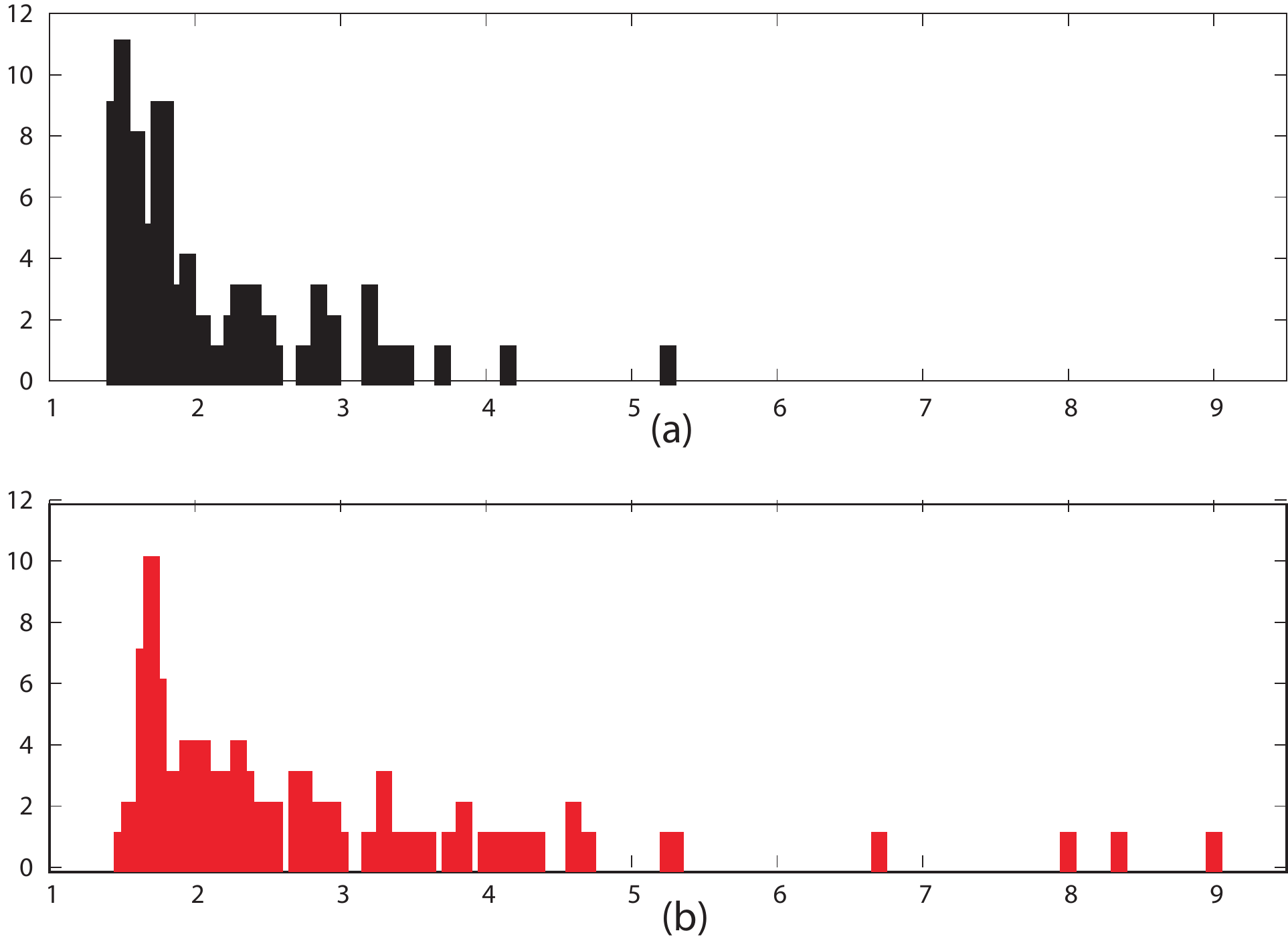}}
\vspace{-2mm}
\caption{{\small (a) $g_3(...)(\Gamma)$ distribution by $100$ RGAPs with 3 anchors in $100\times 100$; 
(b) $g_3(...)(\Gamma)$ distribution by $100$ RGAPs with 3 anchors in $50 \times 100$}}
\label{fig:distribution}
\vspace{-1mm}
\end{figure}

Fig(\ref{fig:distribution})(a)-(b) show the actual 
$g_m(p_1,\cdots,p_m)(\Gamma)$ (average LVT elevation over $\Gamma)$) distribution  
by $100$ RGAPs. In the top graph, $3$ anchors are randomly placed ({\it r.u.p.})
in the entire traversal area while in the bottom graph $3$ anchors 
are  {\it r.u.p.} in the upper half of the traversal area.  
Fig(\ref{fig:distribution})(a) 
clearly exhibit positive skewness. 
This implies that a RGAP over the entire traversal area in general
yields good localization accuracy. Recall that the smaller 
$g_m(p_1,\cdots,p_m)(\Gamma)$ is, 
the better localization is. The $g_m(p_1,\cdots,p_m)(\Gamma)$ 
distribution in Fig(\ref{fig:distribution})(b) 
is more dispersed than that in the Fig(\ref{fig:distribution})(a), 
meaning that 
a RGAP over the half traversal area in general underperforms a RGAP over 
the entire traversal area. 

\begin{table}[hbtp]
\vspace{-1mm}
\centering
\caption{Random Anchor placement}
\vspace{-2mm}
\label{tab:rap}
\begin{tabular}{c|c||c||c||c} \hline
\multicolumn{5}{c}{Random Anchor Placement in The Traversal Area} \\ \hline
\multirow{2}{*}{Anchors} 
& 
\multirow{2}{*}{$\overline{g_m(p_1,..p_m)(\Gamma)}$} 
& 
\multicolumn{1}{|c||}{TPLM} 
& 
\multicolumn{1}{|c||}{LSM} 
& \multicolumn{1}{|c}{GDM} \\ \cline{3-5}
& & ave & ave  & ave  \\ \hline
$3$ & $2.05$ & $0.57$  & $1.52$  & $0.87$  \\ \hline
$4$  & $1.69$ & $0.44$ & $0.79$  & $0.49$  \\ \hline 
$5$  & $1.53$ & $0.40$ & $0.59$  & $0.38$  \\ \hline 
$6$  & $1.48$ & $0.39$ & $0.51$  & $0.33$  \\ \hline 
\multicolumn{5}{c}{Random Anchor Placement in the Half Traversal Area} 
\\ \hline
\multirow{2}{*}{Anchors} 
& 
\multirow{2}{*}{$\overline{g_m(p_1,..p_m)(\Gamma)}$} 
& 
\multicolumn{1}{|c||}{TPLM} 
& 
\multicolumn{1}{|c||}{LSM} 
& \multicolumn{1}{|c}{GDM} \\ \cline{3-5} 
& 
& ave & ave & ave \\ \hline
$3$ & $2.72$ & $0.97$ & $2.82$  & $1.54$  \\ \hline 
$4$ & $1.93$ & $0.52$ & $1.37$  & $0.69$  \\ \hline
$5$ & $1.77$ & $0.46$ & $0.81$  & $0.46$  \\ \hline
$6$ & $1.69$ & $0.44$ & $0.72$  & $0.40$  \\ \hline
\end{tabular}
\vspace{-2mm}
\end{table} 

\begin{figure}[htb]
\centering
%\centerline{\psfig{file=tplm-gdm-lsm-error-dev.eps,height=2.5in,width=3.4in}}
\resizebox{3.4in}{2.5in}{\includegraphics{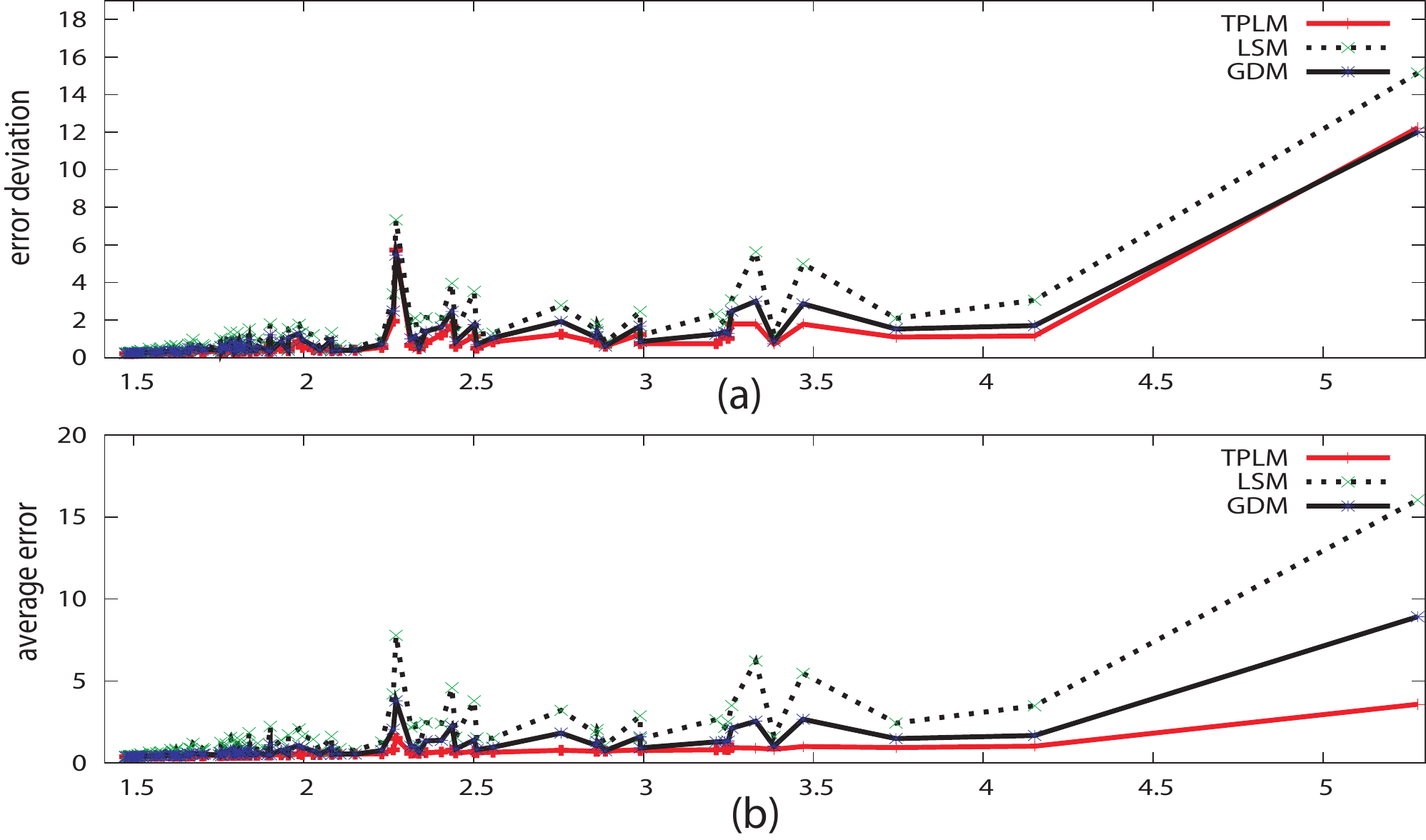}}
\vspace{-2mm}
\caption{{\small (a) error deviation; (b) average error (100 RGAPs with 3 anchors
over the $100\times 100$ region, $\sigma=0.3$)}}
\label{fig:comparisonrandom}
\vspace{-1mm}
\end{figure}
   
Fig(\ref{fig:comparisonrandom}) 
presents the performance curves of LSM, GDM, and TPLM under 
RGAPs, where the 
$x$ axis refers to the 
value of $g_m(p_1,..p_m)(\Gamma)$ and the $y$ axis the error 
deviation and average error.  
As is clear from Fig(\ref{fig:comparisonrandom}), 
TPLM outperforms both LSM and GDM in localization accuracy and reliability: 
both the error deviation and average error curves of TPLM  
are consistently lower than those of LSM and GDM.

Table~\ref{tab:rap} tabulates 
the results of RGAPs with different number of anchors, where 
$\overline{g_m(p_1,..,p_m)(\Gamma)}$ denotes the average 
$g_m(p_1,..p_m)(\Gamma)$ over $100$ RGAPs with a fixed number of anchors.  
It shows that the localization accuracy of TPLM over GDM
is diminished as the number of anchors increases.
It becomes evident that under the noise level of $0.3$, 
the actual performance of LSM, GDM, and TPLM deteriorates 
as the value of $\overline{g_m(p_1,..,p_m)(\Gamma)}$ increases.
This solidifies 
the fact that $g_m(p_1,..p_m)(\Gamma)$ is an effective discriminator
for the anchor placement impact on localization performance. 

\section{Field Experimental Study}
This section focuses on the field test
of a DARPA-sponsored research project,
using the UWB-based ranging
technology from Multispectrum Solutions (MSSI) \cite{Fontana2004}.
and Trimble differential GPS (DGPS). 
Our field testbed area was a $100\times 100$
square meters. It consisted of an outdoor space largely occupied by surface
parking lots and an indoor space inside
a warehouse.
This testbed area was further divided into
$10,000$ non-overlapping $1\times 1$ grids.
Each grid represents the finest positioning resolution to
evaluate RF signal variation.
\begin{table}[hbtp]
\centering
\caption{Field testbed area and anchor placement}
\vspace{-2mm}
\label{tab:fieldtest}
\begin{tabular}{|c|c|c|c|c|} \hline
 $x_0^\prime$ & $y_0^\prime$ & $x_f$ & $y_f$ & $\alpha$ \\ \hline
 $-74.476069$ & $40.537808$ &
$84719$ & $111045$  & $0.381583$\\ \hline 
\multicolumn{5}{|c|}
{Anchor placement} \\ \hline 
\multicolumn{2}{|c|} 
{GPS position} &
\multicolumn{3}{|c|} {transformed position} \\  \hline
$-74.475585$ & $40.538468$ & 
\multicolumn{3}{|c|}{$p_4=(65.345179,52.75145)$}  \\ \hline
 $-74.475287$ & 
$40.53856$ & 
\multicolumn{3}{|c|}{$p_5=(92.580022,52.83239)$}  \\ \hline
$-74.475186$ & 
$40.538294$ & 
\multicolumn{3}{|c|}{$p_6=(89.52274,22.232383)$} \\ \hline
\end{tabular}
\vspace{-3mm}
\end{table}

\begin{figure}[htb]
%\centerline{\psfig{file=cart01-m.eps,height=2in,width=3.4in}}
\resizebox{3.4in}{2in}{\includegraphics{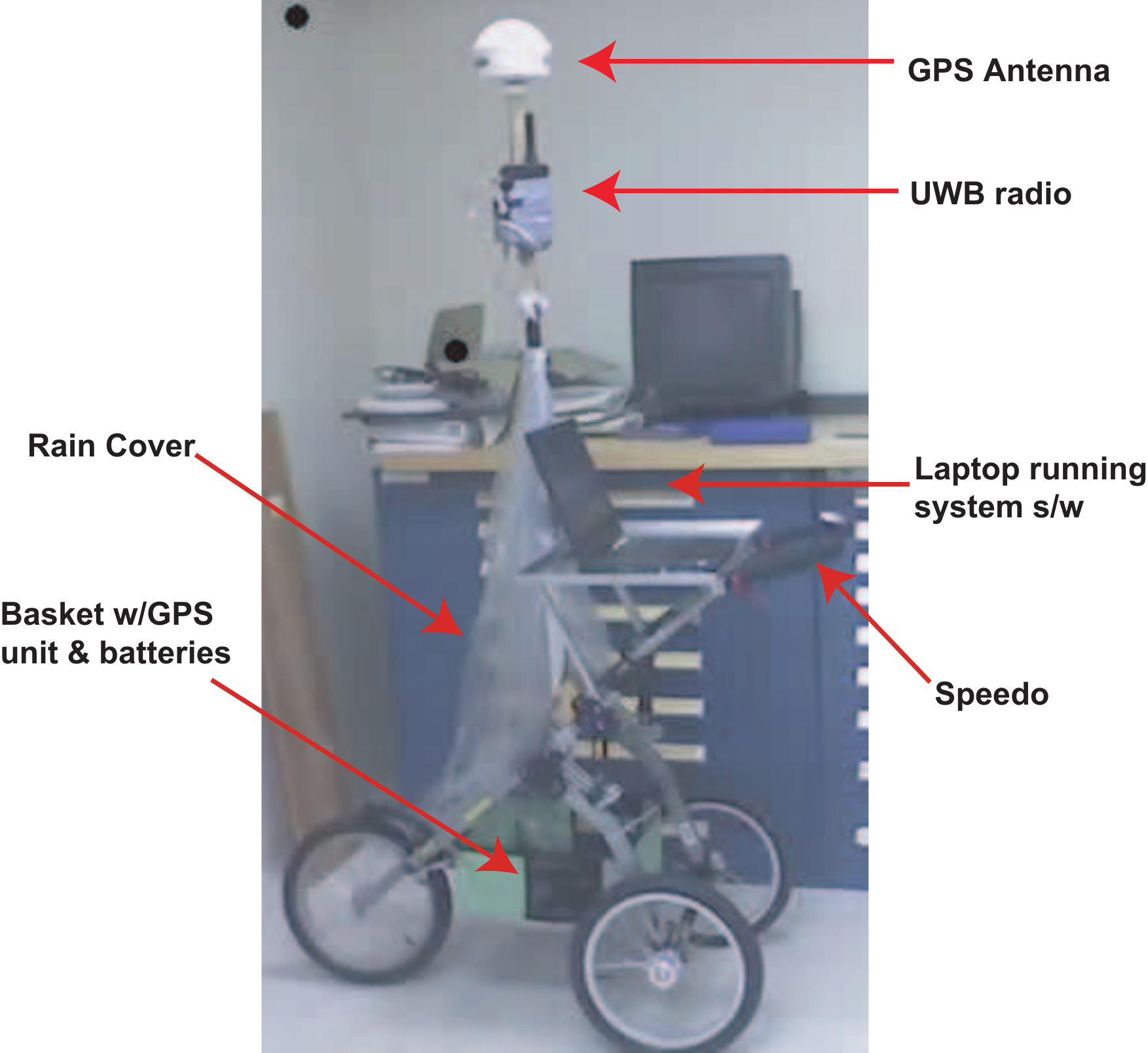}}
\caption{{\small Stroller with a laptop attached with DGPS and UWB ranging devices}}
\vspace{-4mm}
\label{fig:stroller} 
\end{figure}

\begin{figure}[htb]
%\centerline{\psfig{file=ground.eps,height=2.2in,width=3.2in}}
\resizebox{3.2in}{2.5in}{\includegraphics{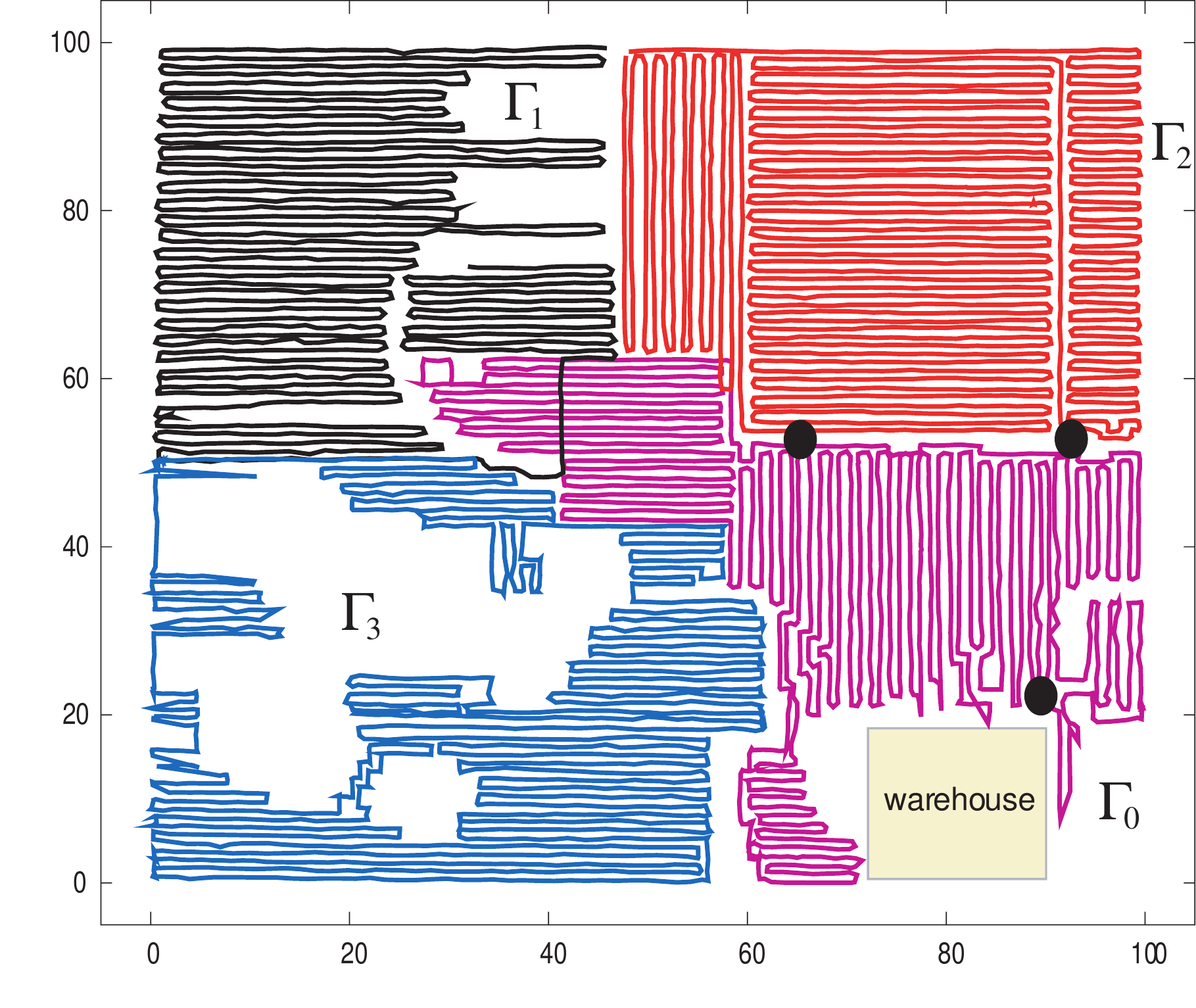}}
\vspace{-2mm}
\caption{{\small Four trajectories ($\Gamma_0,\Gamma_1,\Gamma_2,\Gamma_3$), 
black circles: the positions of anchors, yellow rectangle: warehouse}}
\label{fig:ground}
\vspace{-4mm}
\end{figure}

In the field test, DGPS devices were mainly 
used for outdoor
positioning while MSSI UWB-based devices were used for indoor
positioning. The experimental system was composed of four MNs
equipped with both the Trimble DGPS and MSSI UWB-based ranging devices. Three
UWB devices were used as anchors being placed at known fixed positions. 
Using the known positions of the UWB anchors and real-time distance measurements, 
each MN could establish the current position as well as
that of other MNs locally, at a rate of approximately $2$ samples
per second. Each MN could also establish the position via the DGPS
unit at a rate of $1$ sample per second 
when residing in the outdoor area. The field test area was 
divided into four slightly
overlapping traversal quadrants denoted as $(\Gamma_0,\Gamma_1,\Gamma_2,\Gamma_3)$. 
A laptop (MN) equipped with DGPS and UWB ranging devices was placed in a
modified stroller as shown in Fig(\ref{fig:stroller}), and one 
tester pushed the stroller, traveling along a predefined trajectory inside a quadrant 
(see Fig(\ref{fig:ground})). Each run lasted about $30$ minutes of walk.

\begin{figure*}
\centering
\begin{tabular}{cccc}
\resizebox{1.5in}{1.2in}{\includegraphics{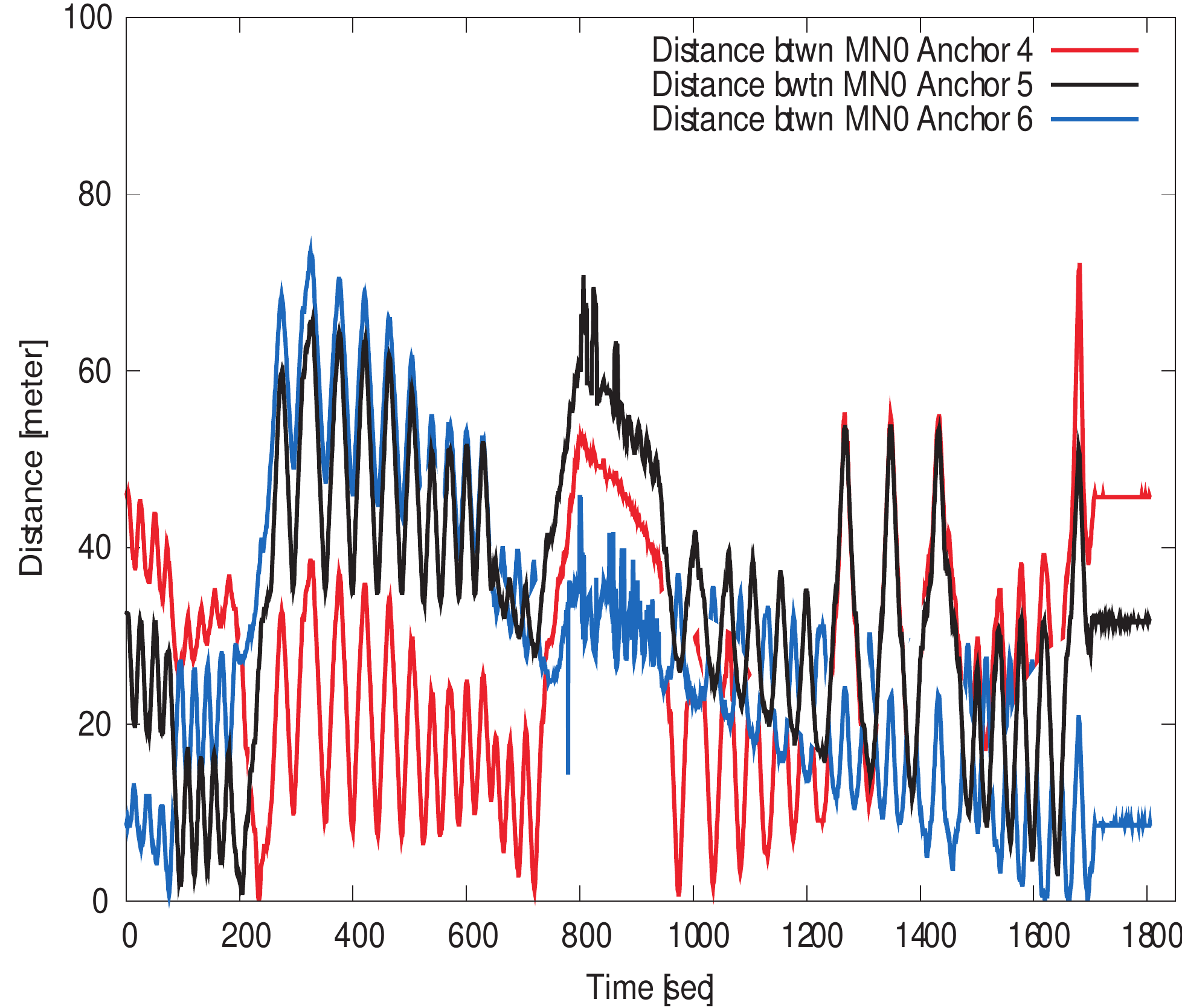}}
& 
\resizebox{1.5in}{1.2in}{\includegraphics{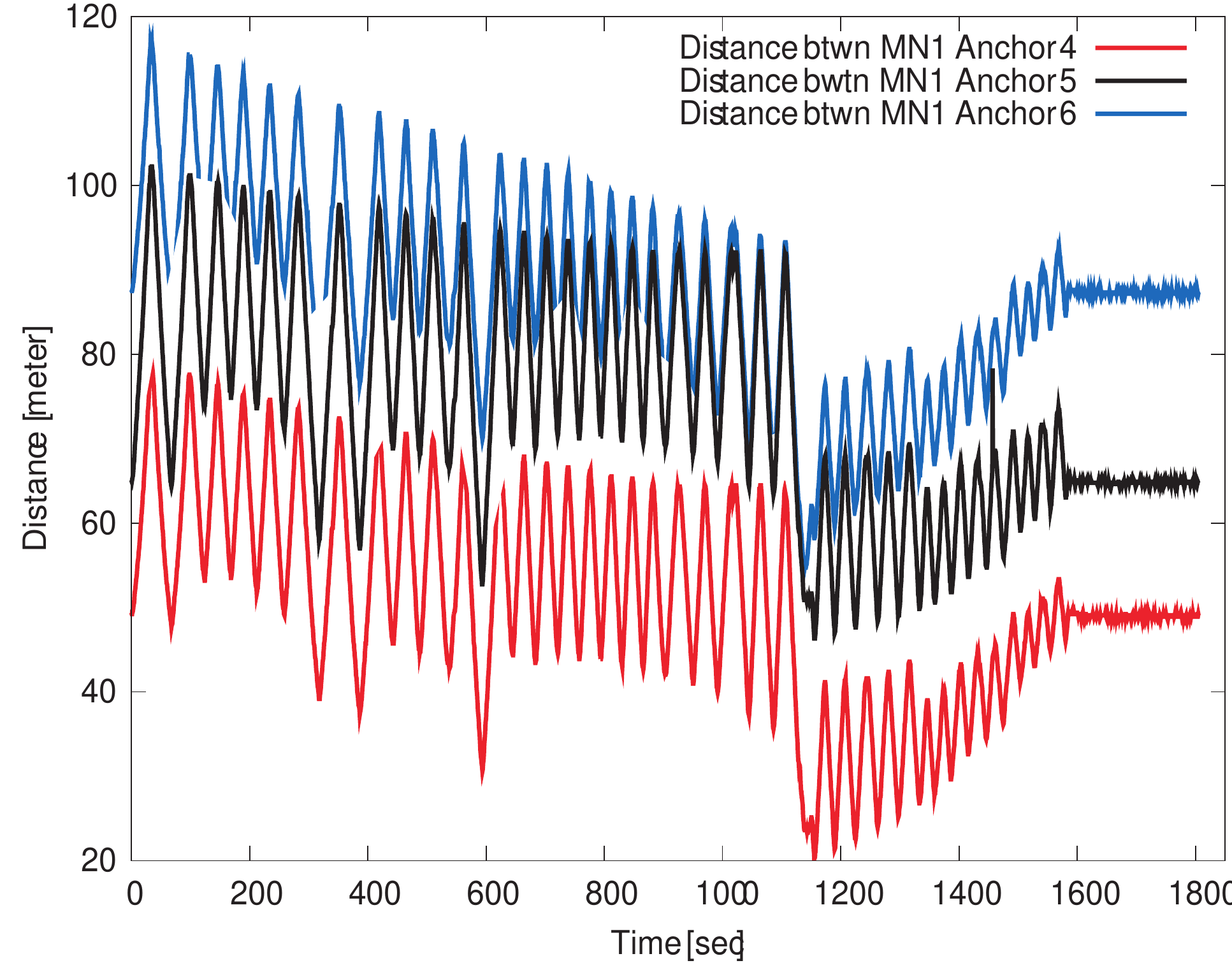}}
& 
\resizebox{1.5in}{1.2in}{\includegraphics{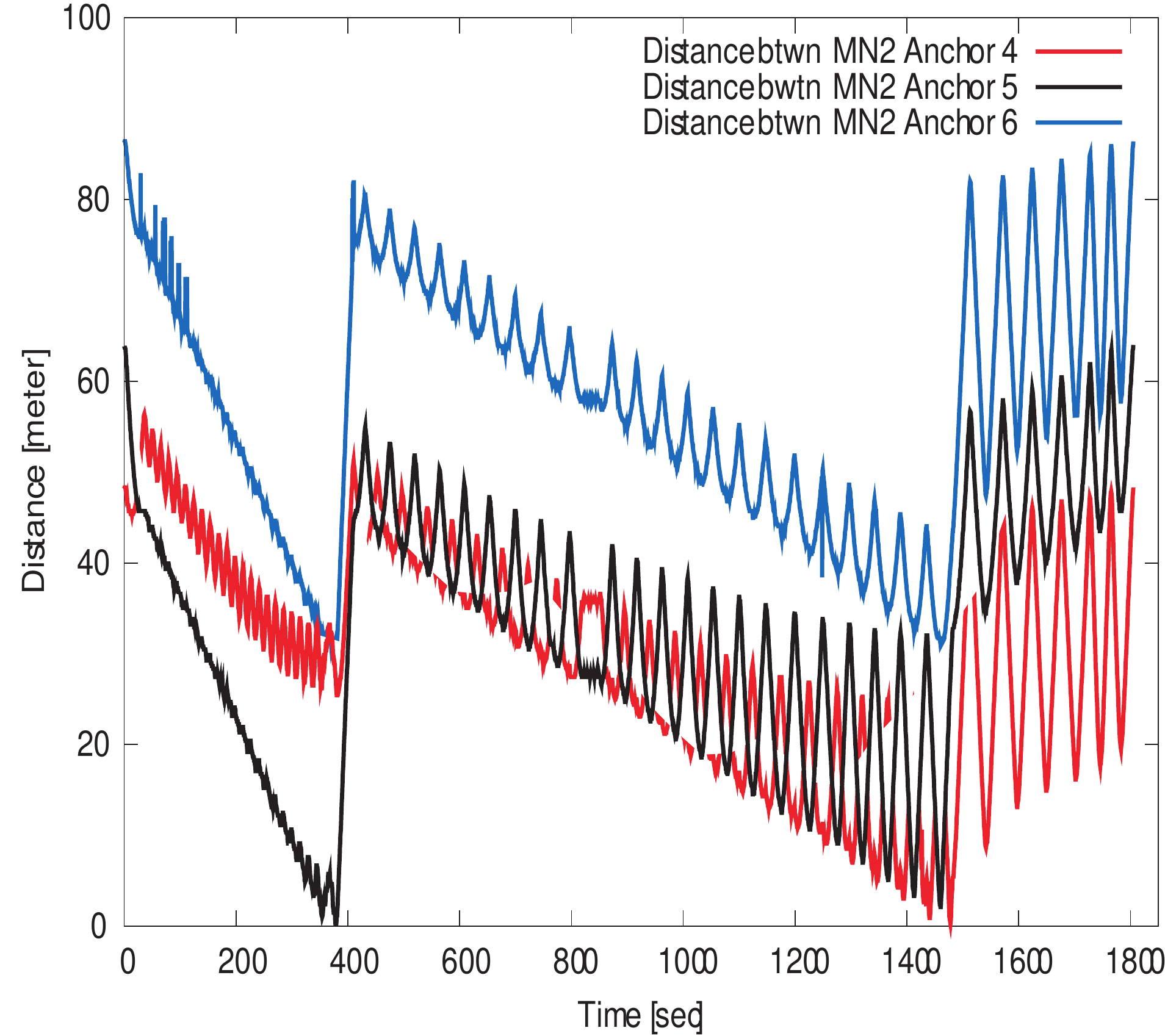}}
&
\resizebox{1.5in}{1.2in}{\includegraphics{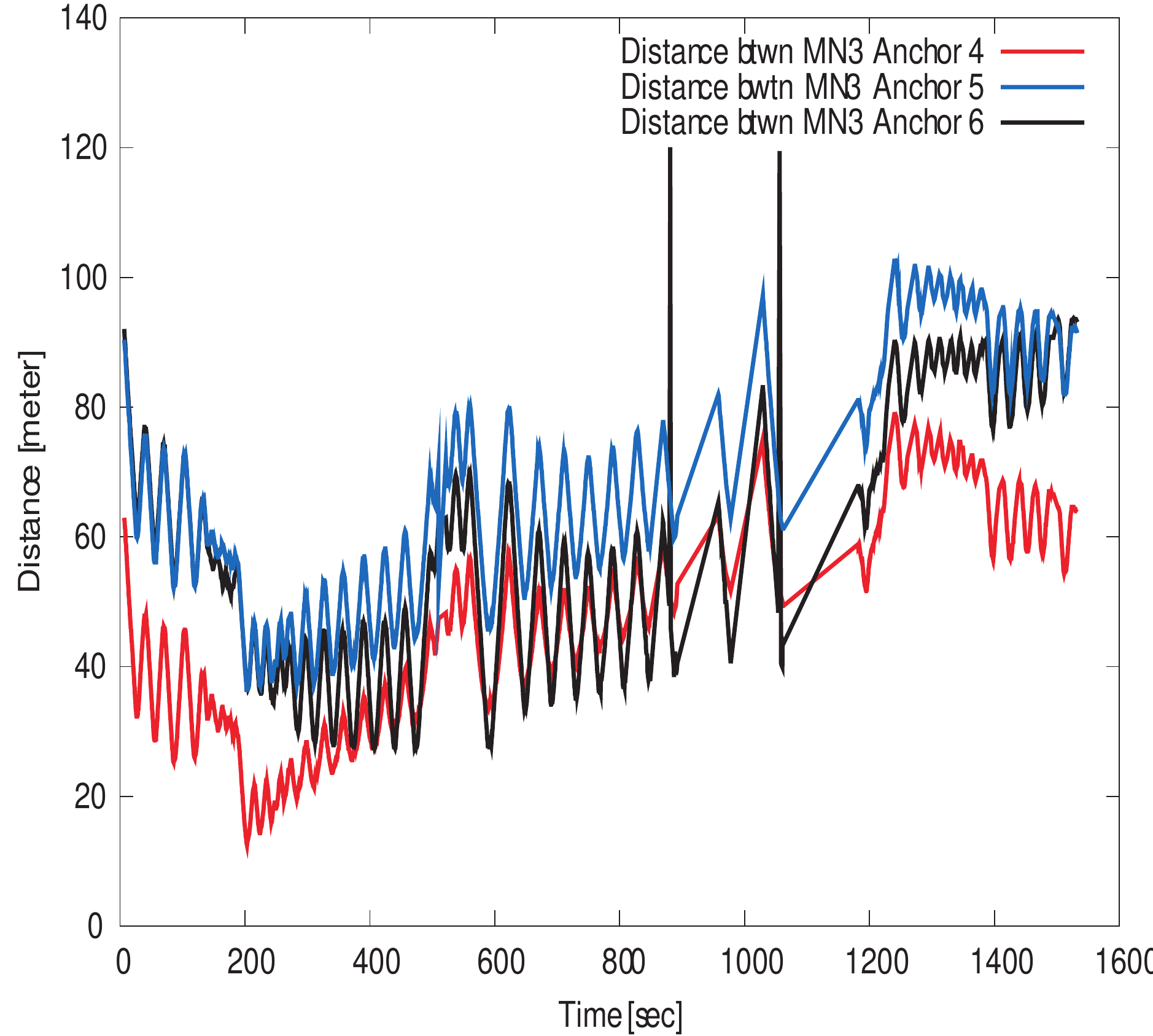}}\\
(a) & (b)& (c) &(d) 
\end{tabular}
\caption{Distance measurements to three UWB anchors: (a) $\Gamma_0$, (b) $\Gamma_1$, (c) $\Gamma_2$, (d) $\Gamma_3$}
\vspace{-3mm}
\label{fig:mn0-mn3} 
\end{figure*}

With repeated trial runs, we found that the UWB signal could barely penetrate 
one cement wall of the warehouse. To
establish positions inside the warehouse, we 
placed one anchor close
to the main entrance of the warehouse while 
placing two other anchors around the center of the field testbed area. 
Fig(\ref{fig:ground}) showed the four trajectories 
($\Gamma_0,\Gamma_1,\Gamma_2,\Gamma_3$) in the field test, which 
are translated from the GPS trajectories.
A part of $\Gamma_0$ trajectory was inside the warehouse, thus the part of GPS of 
$\Gamma_0$ was not available. 
Table\,\ref{tab:field-ground} provides the calculated results of 
$g_3(p_4,p_5,p_6)(\Gamma)$ 
using the GPS trajectory data, indicating 
that $\Gamma_0$ would yield the most accurate 
localization while $\Gamma_1$ produced
the least accurate localization.
\begin{table}[htbp]
\centering
\caption{$g_3(p_4,p_5,p_6)(\Gamma)$ of trajectories in field testbed}
\vspace{-2mm}
\label{tab:field-ground}
\begin{tabular}{c|c} \hline  
$g_3(p_4,p_5,p_6)(\Gamma_0)$
& 
$g_3(p_4,p_5,p_6)(\Gamma_1)$  \\ \hline
$1.667025$ & $1.967605$  \\ \hline 
$g_3(p_4,p_5,p_6)(\Gamma_2)$
& $g_3(p_4,p_5,p_6)(\Gamma_3)$ \\ \hline
$1.787370$ & $1.835048$ \\ \hline
\end{tabular}
\end{table} 

\begin{figure}
\centering
%\centerline{\psfig{file=tplm-gdm-lsm-scale.eps,width=3.5in,height=2.5in}}
\resizebox{3.5in}{2.5in}{\includegraphics{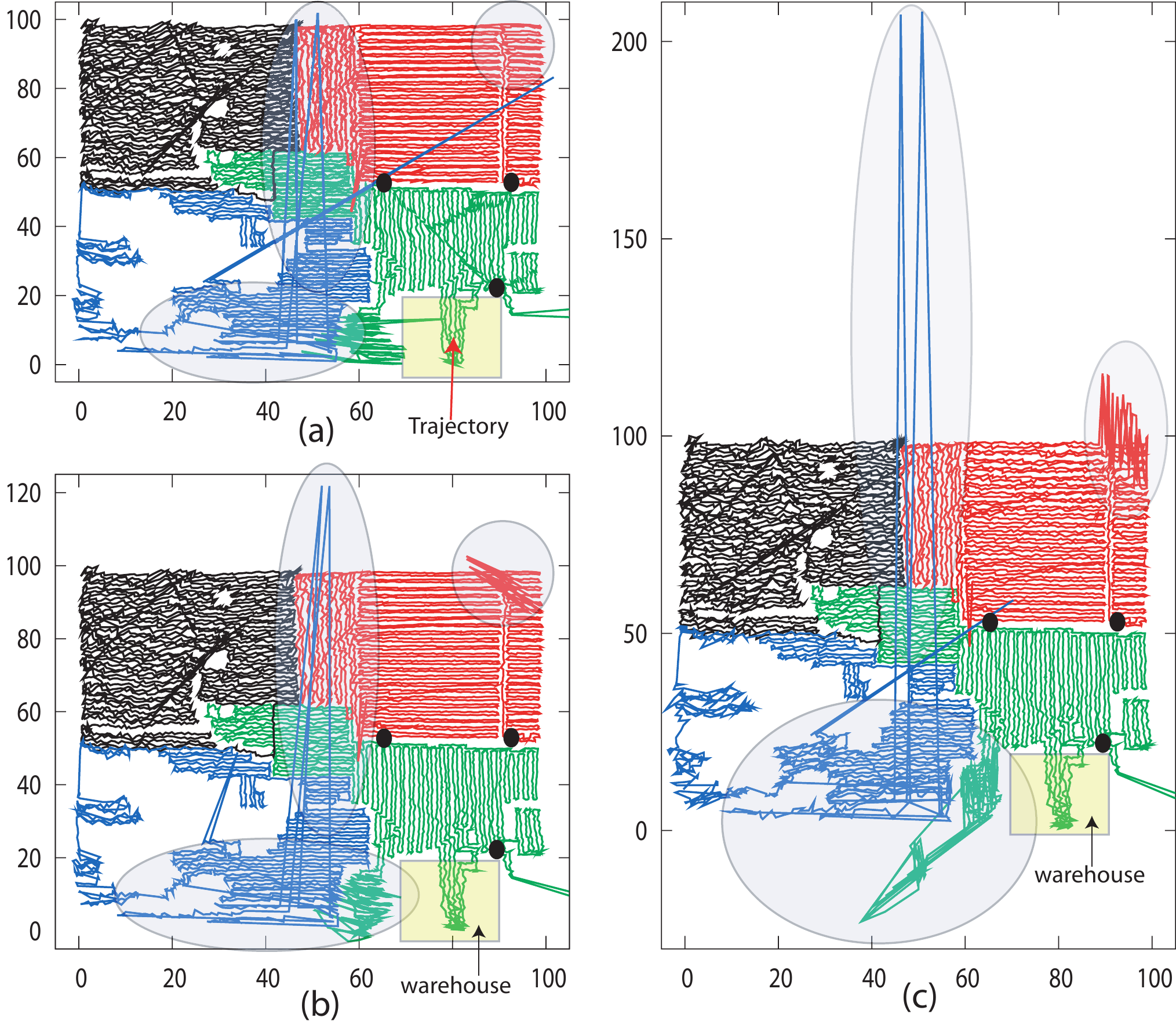}}
\vspace{-2mm}
\caption{{\small Restored Trajectories by (a) TPLM, (b) GDM, (c) LSM (black circles: positions of 
UWB anchors)}}
\label{fig:tplm-gdm-lsmrestore}
\vspace{-1mm}
\end{figure}
 
In the field testbed, the trajectory of each MN was
controlled by an individual tester during  
a $30$-minute walk. For purpose of the primary experiment being
conducted, the strollers were fitted with bicycle speedometers to allow
the tester to control his speed, in order to produce sufficient reproducibility
for the primary experiment, but not for our testing of localization.
Due to inherent variability in each individual movement, 
an objective assessment of 
localization accuracy without a 
ground truth reference
is almost impossible. 
For a performance comparison, we used GPS trajectories as a 
reference for visual inspection 
of the restored trajectories by LSM, GDM, and TPLM.

Three curves in Fig(\ref{fig:mn0-mn3})(a)-(d) 
represent the field distance measurements between a tester and the three UWB anchor devices 
during a $30$-minute walk in trajectories $(\Gamma_0,\Gamma_1,\Gamma_2,\Gamma_3)$. 
it is fairly obvious that measurement noise in different 
trajectories vary widely: The distance measurement curves in trajectories $(\Gamma_0,\Gamma_3)$
are discontinuous and jumpy in Figs(\ref{fig:mn0-mn3})(a)\&(d),  in contrast to 
the relatively smooth  
distance curves in trajectories $(\Gamma_1,\Gamma_2)$ 
in Fig(\ref{fig:mn0-mn3})(b)\&(c). Such a discontinuity in measurements 
occurred when testers were traveling in an area 
where UWB devices on stroller had no direct line of sight to UWB anchors, 
thus introducing 
additional noise in $\Gamma_0, \Gamma_3$.

Figs(\ref{fig:tplm-gdm-lsmrestore})(a)-(c) 
show the restored trajectories by TPLM, GDM, and LSM using 
the UWB ranging technology in the field test.
A visual inspection suggests that 
LSM was extremely prone to noise as
its restored trajectories 
were appreciably distorted beyond recognition in some parts. 
As indicated in Fig(\ref{fig:tplm-gdm-lsmrestore})(b), GDM gave an obvious error reduction over
LSM but at the expense of computational cost. 
The most visually perceived difference between LSM and GDM can be 
seen in the circled areas in Figs(\ref{fig:tplm-gdm-lsmrestore})(b)-(c). 
By contrast, the difference between GDM
and TPLM can be visualized in the circled areas
Figs(\ref{fig:tplm-gdm-lsmrestore})(a)-(b) where TPLM 
produced a detail-preserved 
but slightly distorted contour of the trajectory. A further offline analysis showed
that on average GDM takes $18.87ms$ per position establishment,
while TPLM/LSM take $0.4/0.35ms$.
 
\section{Conclusion}
This paper studies
the geometric effect of anchor placement
on localization performance.
The proposed approach allows the construction of 
the least vulnerability tomography (LVT) for comparing the geometric impact
of different anchor placements. 
As a byproduct, we propose a two-phase localization method.

To validate theoretical results, we conduct comprehensive simulation experiments
based on randomly generated anchor placements and different noise models. 
The experimental results agree with our anchor placement impact analysis.   
In addition, we show that TPLM outperforms LSM by a huge margin in 
accuracy. TPLM performs much faster than GDM and 
slightly better than GDM in localization accuracy.
The field study shows that TPLM is more robust against 
noise than LSM and GDM. 

In this paper, we adopt a widely used assumption that
measurement noise is independent of distance. This assumption
is somewhat unrealistic in some practical settings. 
In our future research, we will perform experiments using 
UWB ranging technology to study the noise-distance relation
in indoor and outdoor environments. The anchor placement impact
analysis will be refined to incorporate more realistic noise-distance 
models.

\bibliographystyle{plain}
\bibliography{position}

\end{document}